\documentclass[10pt,twoside]{article}

\usepackage{times}
\usepackage{bbm}
\usepackage{amsmath}
\usepackage{graphicx}

\usepackage[all]{xy}

\newcommand{\cc}{\mathbbm{C}}
\newcommand{\rr}{\mathbbm{R}}
\newcommand{\zz}{\mathbbm{Z}}
\newcommand{\nn}{\mathbbm{N}}
\newcommand{\id}{\mathbbm{I}}

\newcommand{\dist}{{\rm dist}}
\newcommand{\tr}{{\rm tr}}

\newcommand{\xminid}[1]{
	\xymatrix@C=5mm{#1}
}
\newcommand{\Par}{\ar@{-}[r]}
\newcommand{\Pau}{\ar@{-}[u]}
\newcommand{\Pad}{\ar@{-}[d]}
\newcommand{\Pal}{\ar@{-}[l]}

\begin{document}
{\LARGE{\bf Entanglement and tensor network states}}

\medskip
{Jens Eisert}

{\it Dahlem Center for Complex Quantum Systems}

{\it Freie Universit{\"a}t Berlin, 14195 Berlin, Germany}

\tableofcontents

\newpage
\section{Correlations and entanglement in quantum many-body systems}

\subsection{Quantum many-body systems}
In this book chapter we will consider {\it quantum lattice systems} as they 
are ubiquitous in the condensed matter context or in  situations that mimic condensed matter systems, as provided, say, 
by systems of cold atoms in optical lattices. 
What we mean by a quantum lattice system is the following: We think that we
have an underlying lattice structure given: some lattice that can be captured
by a graph. The vertices of this graph are associated with a quantum degree of freedom each, referred to as {\it constituents},
while edges correspond to neighbourhood relations. Interactions in the physical system are usually {\it local}, which means that all
constituents only directly interact with finitely many neighbours on the lattice.
Particularly important is the situation when all interactions except from direct nearest neighbour interactions can be safely neglected. 
Quantum lattice models of this type
capture strongly correlated materials often exhibiting interesting electronic and magnetic properties. They  serve as theoretical
laboratories allowing to study features of topological order and non-conventional phase transitions. 
Quantum systems well modelled by lattice models in this sense also show a wealth of phenomenology in out-of-equilibrium
situations, to mention only a few reasons why this kind of physical system is interesting.

In this book chapter, 
we will provide a brief introduction into {\it tensor network approaches to the study of quantum lattice models}. 
The position taken may be slightly unusual in the sense that a
rather strong emphasis is put onto methods and ideas of description, and not so much on the phenomenology itself (which can then be derived from such a
description, needless to say). Given that it is the very development of the
toolbox of tensor network methods itself that is being reviewed here, one that has led to many recent new insights,
this seems a healthy standpoint. 

But there is yet another shift of emphasis that may be somewhat
 unexpected: namely that rather quantum states and not so much Hamiltonians are in the focus of attention. Here
 it is mostly the very nature of ground and thermal states itself that is being considered and studied,
while Hamiltonians primarily reenter via the concept of a parent Hamiltonian. The main message of this book chapter can be summarised in a single paragraph:
\smallskip

 {\it Many natural quantum lattice models have ground states that are little, in fact very little, entangled in a precise sense. This shows that `nature is lurking in 
 some small corner of Hilbert space', one that can be essentially efficiently parametrized. This basic yet fundamental insight allows for a plethora of new methods for the
 numerical simulation of quantum lattice models using tensor network states, 
 as well as a novel toolbox to analytically study such systems.}\footnote{In this book chapter, we will discuss the 
 elements of this framework, while at the same time we cannot provide a comprehensive review. This 
 chapter will still contain slightly more material than what is covered in the course. For recent reviews covering related topics, 
 see Refs.\ \cite{SchollwoeckAge,AreaReview,VerstraeteBig,SchuchChapter,OrusChapter}.}
 
\subsubsection{Quantum lattice models in condensed-matter physics}

We start by discussing the concept of a quantum lattice model.\index{Quantum lattice model}
The underlying {\it graph $G=(V,E)$} 
capturing the {\it lattice} may, in principle, be any graph, where $V$ is the vertex and $E$ the edge set. $\dist(.,.)$ is then the graph-theoretical distance, so the 
minimum number of steps one has to walk on the graph in order to get from one vertex to another.
Particularly important are, however, regular graphs, and even more so particularly common regular graphs. In fact, most of the time we will be concerned with
simple {\it cubic lattices} $V=L^{\cal D}$ in dimension ${\cal D}$, and specifically one-dimensional lines for which ${\cal D}=1$.
But also other lattices such as {\it triangular lattices} or {\it Kagome lattices} often emerge naturally. 
$n=|V|$ is referred to as the {\it system size}. The quantum degree of freedom at each vertex can be a spin system of dimension $d$ -- which 
will be the situation in the focus of attention in this chapter -- 
or a {\it bosonic} or a {\it fermionic} degree of freedom (which we touch upon). 
The entire Hilbert space of the system is hence given by ${\cal H}= (\cc^d)^{\otimes n}$ in case of spin models.
For bosons and fermions we consider the {\it Fock space}. Say, if ${\cal K}$ is the Hilbert space associated with a single fermion, then we make use of the Fock space
${\cal F}= \wedge^\ast ({\cal K})$. Clearly, the dimension of the Hilbert space grows exponentially with the system size, specifically $\dim({\cal H})=\Omega(d^n)$ for
a spin model,
which means that a numerical exact diagonalisation of the underlying Hamiltonian is doomed to failure for already moderately large system sizes $n$. In fact, a naive
diagonalisation of the Hamiltonian without exploiting any additional structure would require $O(d^{3n})$ operations, 
clearly infeasible for large quantum systems.

\subsubsection{Local Hamiltonians}

All Hamiltonians considered here will feature finite-ranged interactions, which means that they can be written in the form
\begin{equation}
	H= \sum_{j\in V}h_j,
\end{equation}		
where each $h_j$ is non-trivially supported only on finitely many sites in $V$ (but not necessarily on site $j$ only). 
It is called {\it $k$-local} if each $h_j$ is supported on at most $k$ sites,
and {\it geometrically $k$-local} if each $h_j$ is supported on a $V_j$ with
$\max_{a,b\in V_j} \dist (a,b)= k-1$. This is a most natural situation: Each constituent interacts then merely with its immediate neighbours in the lattice.
We will restrict attention to such geometrically $k$-local Hamiltonians, in fact with few exceptions to nearest-neighbour models. One also often writes $\langle j,k\rangle$
for neighbours, so for sites $j,k\in V$ such that $\dist(j,k)=1$. Similarly, one also calls any observable that is non-trivially supported only on neighbouring sites a {\it local
observable}.

There are a number of famous examples of such local Hamiltonians. A Hamiltonian that has specifically been studied countlessly many times -- for good reasons --
is the XY-model Hamiltonian on a one-dimensional line with $n$ sites \cite{LSM}, where $V=\{1,\dots, n\}$ and ${\cal H}=(\cc^2)^{\otimes n}$,
\begin{equation}
	H = -\frac{1}{2}\sum_{\langle j,k\rangle}
	\left(
	\frac{1+\gamma}{4}X^{(j)} X^{(k)}  + \frac{1-\gamma}{4} Y^{(j)}  Y^{(k)} 
	\right) - \frac{\lambda}{2}\sum_{j\in V} Z^{(j)} ,
\end{equation}
where $\gamma\in \rr$ is called the {\it anisotropy parameter} and $\lambda\in \rr$ is the {\it magnetic field}. The matrices
\begin{equation}
	X = \left[
	\begin{array}{cc}
	0 & 1\\
	1 & 0
	\end{array}
	\right],\,\,
	Y = \left[
	\begin{array}{cc}
	0 & -\i\\
	\i & 0
	\end{array}
	\right],\,\,
	Z = \left[
	\begin{array}{cc}
	1 & 0\\
	0 & -1
	\end{array}
	\right]
\end{equation}
are the familiar {\it Pauli matrices}, the index referring to the site they are supported on. It should be clear that this Hamiltonian is a local Hamiltonian of the above form.
Simple as this model is, it can be easily analytically solved by considering it as a problem of free fermions. It already features notions of criticality, and is indeed often
studied as a paradigmatic lattice model -- not the least because everything imaginable can be said about it. The situation of $\gamma=0$ is particularly important; then the 
model is also called {\it isotropic XY model} or {\it XX model}.

\subsubsection{Boundary conditions}

Depending on whether one identifies for the one-dimensional line $V=\{1,\dots, n\}$ the
site $n+1$ with the site $1$ or not, one says that says that one has {\it open} or {\it periodic} boundary conditions. We will also consider open and periodic boundary conditions
for other cubic lattices, where $V\in L^{\cal D}$. For periodic boundary conditions, one encounters then the topology of a torus in ${\cal D}$ dimensions.

\subsubsection{Ground states, thermal states, and spectral gaps}

The lowest energy eigenvectors of the Hamiltonian, so the normalised state vectors that minimise $\langle \psi| H |\psi\rangle$, 
form a Hilbert space ${\cal G}$, the {\it ground space}, a subspace of ${\cal H}$; one also often refers to the {\it ground state manifold}. If the ground space is one-dimensional, the ground
state is {\it unique}, otherwise it is called {\it degenerate}. Ground states often capture the low temperature physics and their study is ubiquitous in theoretical
physics. The energy associated with the ground space is the {\it ground state energy}, usually denoted as $E_0$. {\it Ground state expectation values} 
will be denoted as $\langle O\rangle$, $O\in {\cal B}({\cal H})$ being some observable. Particularly important are  {\it local observables} $O_A$
which are supported on finitely many sites $A\subset V$ only (actually most prominently on just a single site).


The {\it Hamiltonian gap} is the energy gap from the ground space to the first excited state, so
\begin{equation}
	\Delta E = \inf_{|\psi\rangle \in {\cal H}\backslash {\cal G}}\langle \psi| H |\psi\rangle - E_0.
\end{equation}
If $\Delta E=0$ for a family of Hamiltonians in the thermodynamic limit of $n\rightarrow\infty$, then one says that the system is {\it gapless} or {\it critical}.
Such critical models can be beautifully captured in the framework of {\it conformal field theory} \cite{CFT}
which is outside the scope of this book chapter.
If a positive gap exists in the thermodynamic limit, it is {\it gapped}. 

\subsection{Clustering of correlations}

Since direct interactions are local and constituents directly see their immediate neighbours only, one should expect that correlation functions between different constituents somehow
decay with the distance in the lattice. The correlation functions hence should be expected to inherit the locality of interactions. It turns out that this is indeed the case. 

\subsubsection{Clustering of correlations in gapped models and correlation length}
Specifically,
for gapped models, correlation functions always decay exponentially with the distance. This effect is also called {\it clustering of correlations}. Nearby lattice sites will still be 
correlated to some extent, but these correlations become negligible for large distances. So if for a family of models $\Delta E>0$ (and under very mild conditions on the lattice
$G$ which are always satisfied for natural finite-dimensional lattices), then \index{Clustering of correlations} \cite{Decay}
\begin{equation}\label{Correla}
	\left|
	\langle O_A O_B\rangle -\langle O_A\rangle\langle O_B\rangle 
	\right|\leq C e^{-\dist(A,B) \Delta E/(2v)} \| O_A\|\, \|O_B\|,
\end{equation}
for some suitable constants $C,v>0$.
The length scale 
\begin{equation}\label{CorrelationLength}	
	\xi	:= \frac{2v}{\Delta E}>0
\end{equation}	
emerging here is the {\it correlation length}: it is the characteristic length scale on which correlations disappear.
The feature of clustering of correlations has long been suspected to be generically valid in gapped models and has been `known' for a long time. A rigorous proof of this can be 
obtained in a quite beautiful way using {\it Lieb-Robinson bounds} \cite{LR}, which are bounds to the speed of information propagation in time in quantum lattice models
with local Hamiltonians, in terms of the {\it Lieb-Robinson velocity} $v>0$. 
By using a clever representation in the complex plane, one can relate this statement --  which as such relates to dynamical features -- to 
ground state properties \cite{Decay}.  $\|.\|$ in the above expression 
is the {\it operator norm}, so the largest singular value: It grasps the `strength' of the observable.
Again, if $A$ and $B$ are far away in gapped models, in that $\dist(A,B)$ are much larger than the correlation length, 
then the correlation function will essentially disappear.

\subsubsection{Algebraic decay in gapless models}

For gapless models the above is no longer true. Generically, correlation functions of gapless models decay algebraically with the distance in the lattice.
Then there is no longer a length scale associated with the decay of correlation functions. {\it Conformal field theory} \cite{CFT}
provides a framework of systems exhibiting a 
conformal symmetry, which can be applied to the study of critical quantum lattice systems.

\subsection{Entanglement in ground states and area laws}

Yet, there is a stronger form of locality inherited by the locality of interactions than merely the decay of two-point correlation functions between any two sites.
These stronger forms of locality are captured using concepts of entanglement in quantum many-body systems. Indeed, the insight that concepts and methods of {\it
entanglement theory} -- as they originally emerged in quantum information theory -- can be applied to the study of quantum many-body system triggered an explosion
of interest in the community. The significance of this for grasping quantum many-body systems in condensed-matter physics with tensor networks will become clear in a minute.

\subsubsection{Entanglement entropies}

Let us imagine we have a gapped lattice model prepared in the ground state, based on some
lattice $G=(V,E)$. We now single out a certain subset $A\subset V$ of sites, some region,  
and consider its complement $B:= V\backslash A$. 
This subset will be associated with a reduced state $\rho_A=\tr_B (\rho)$. The reduced state
alone will allow us to compute every expectation value of observables supported on $A$; it is
obtained by fixing an orthonormal basis in $B$ and taking the
partial trace. Now, what will the von-Neumann entropy \index{Entanglement entropy}
\begin{equation}
	S(\rho_A) = -\tr(\rho_A\log_2 \rho_A)
\end{equation}
of the state $\rho_A$ be? Of course, the entropy of the entire ground state $\rho$ will vanish, so $S(\rho)=0$, it being a pure state,
but this is not true for the entropy of reduced states.
If the ground state is unique, so if it is a pure state, which we are assuming here, this entropy reflects the {\it degree of entanglement} \cite{Bennett}
of the system $A$ with respect to its complement. If $A$ and $B$ are in a product state and no entanglement is present, then $S(\rho_A)=0$. Otherwise, the entropy
will be larger the more entangled the sub-systems are, being bounded from above by the maximum value  $S(\rho_A)\leq |A|\log_2(d)$.

 Quantum correlations make the entropy of reduced states become non-vanishing. In fact, according to a meaningful
axiomatic quantification of asymptotic entanglement manipulation, the von-Neumann entropy {\it uniquely} quantifies the entanglement content \cite{EntanglementUniqueness}
in a sense. 
This {\it entanglement measure}
is called {\it entropy of entanglement} or {\it entanglement entropy}. Note that this is only a meaningful measure of entanglement, an entanglement monotone, as one says, for
pure states; so in our context, if the ground state is non-degenerate. For mixed states, one can still compute the entropy of the reduced state, but this would no longer amount to
an entanglement monotone (and this quantity then no longer captures the entanglement in the ground state, but has to be replaced by other measures that we will discuss below).

\subsubsection{Area laws for the entanglement entropy}

So how does $S(\rho_A)$ scale with the size $|A|$ of the region $A$? Detailed knowledge about the entropy will here be less important than the general scaling behavior
in the asymptotic limit of large regions. Questions of a similar type have a long tradition and were first asked in the context of the scaling of black hole entropies \cite{Srednicki}.
Naively, but at the same time naturally, one might expect this quantity to scale {\it extensively} with the size $|A|$: 
This would mean that $S(\rho_A)= O(|A|)$. After all, entropies of Gibbs states in statistical mechanics are known to scale extensively,
so one might think that the same intuition may be applied here. And indeed, for `generic states', so for random states, this is true with overwhelming probability. One can rigorously
define such random vectors using the Haar measure of the unitaries acting on ${\cal H}$, and finds that the expected entanglement entropy indeed takes the maximum value 
$|A|\log_2(d)$, up to a correction
that is exponentially small in the size of $B$.

This intuition, however, is not correct. Instead, one finds that the entropy scales as the {\it boundary area} of $A$, so\index{Area law}
\begin{equation}
	S(\rho_A) = O(| \partial A |).
\end{equation}	
One then also says that the entanglement entropy satisfies an {\it area law for the entanglement entropy}.
This boundary $\partial A$ of the region $A$ is defined as
\begin{equation}
	\partial A:= \left\{
	j\in A: \exists k\in B\,\text{with} \,
	\dist(j,k)=1
	\right\},
\end{equation}
\begin{figure}[h!]
 \centering
\includegraphics[width=0.47\textwidth]{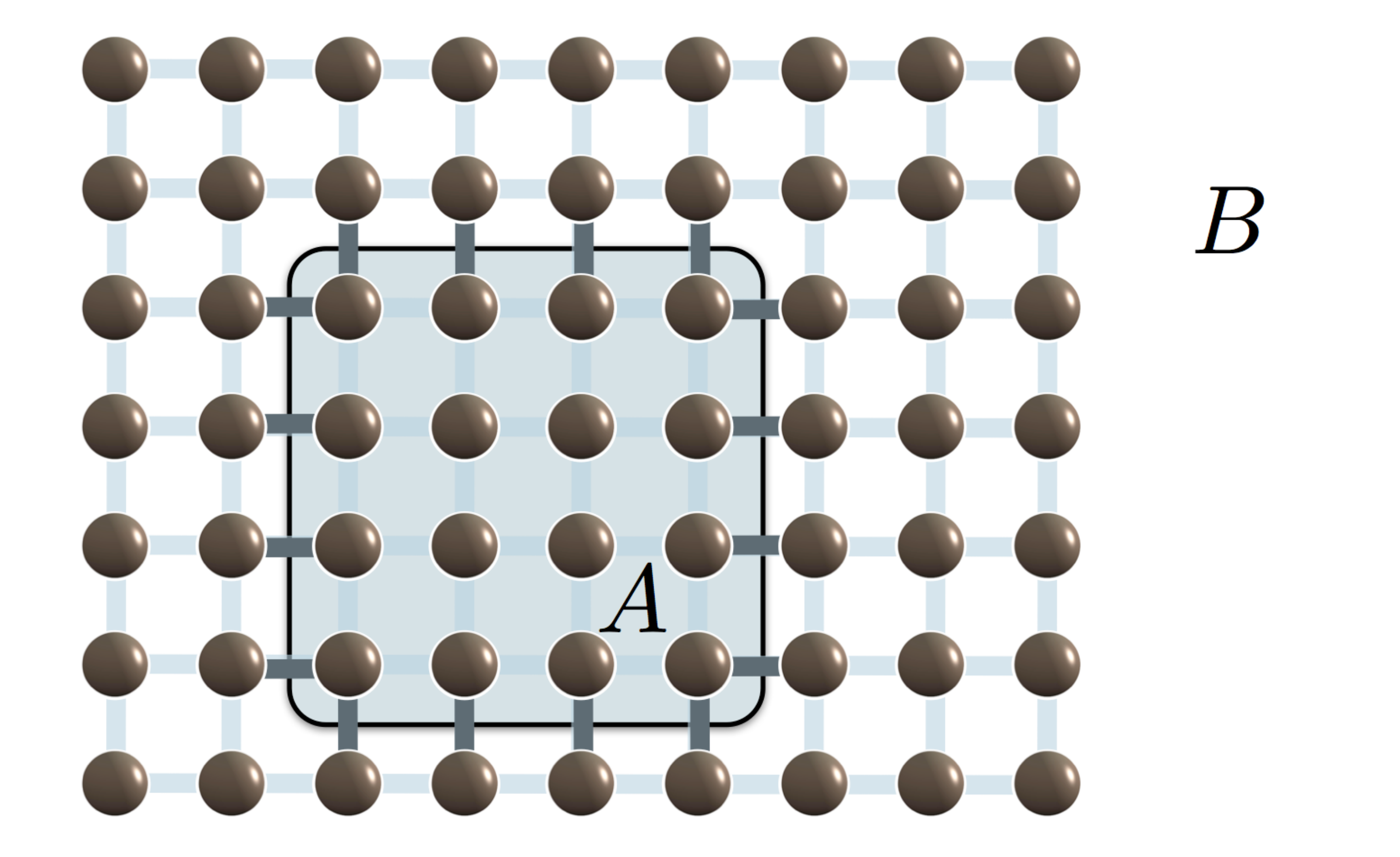}.
\end{figure}

For a one-dimensional system, this boundary consists of two sites only, in a cubic lattice of dimension ${\cal D}>1$ it contains $O(L^{{\cal D}-1})$ many sites. 
That is, ground states of gapped models are less entangled than they actually could be, in fact {\it much less} entangled. This will be key to the understanding for what
follows. For a comprehensive review on area laws, see Ref.\ \cite{AreaReview}.

\subsubsection{Proven instances of area laws}

Such an area law has been proven to be valid for a number of cases:
\begin{itemize}
\item For any gapped one-dimensional system with a unique ground state \cite{OneD}. The proof again relies on Lieb-Robinson bounds mentioned before, albeit used
in a much more
sophisticated way than in order to show the clustering of correlations.
This proof has in the meantime been significantly tightened using ideas of the {\it detectability lemma} \cite{Detectability}.

\item For gapped {\it free bosonic and fermionic models}, so for models where the Hamiltonian can be written as a quadratic polynomial of bosonic or fermionic annihilation and creation operators, the area law is known in fact for arbitrary lattices in any dimension \cite{Area,Area2}.

\item For free bosonic models, the area law is even known to hold for critical, gapless models for cubic lattices of dimension ${\cal D}>1$ \cite{Halfspace}.

\item For some classes of states such as instances of so-called {\it graph states} \cite{Graphs} the area law holds by construction.
\item Most importantly for the purposes of these lecture notes, {\it matrix product states} in one-dimensional systems and {\it projected entangled pair} states for higher-dimensional
systems also satisfy an area law \cite{PEPSArea}. As we will see in a minute, this insight is not a detail at all: It is at the very 
heart of the insight why gapped quantum many-body systems can actually be numerically simulated using tensor network states.

\item Once one has identified a system that satisfies an area law in higher-dimensions, one still obtains an area law for local Hamiltonians that are in the same quantum phase. This has been shown by making use of ideas of {\it quasi-adiabatic evolution}, 
Lieb-Robinson bounds and bounds to the generation of entanglement via local dynamics 
 \cite{VerstraetePerturbation,SpirosPerturbation}.
 \item A different route is taken in Ref.\ \cite{BrandaoAreaLaw}, 
 where it is shown that the exponential decay of correlations of the type as in Eq.\ (\ref{Correla})
 implies an area law for the entanglement entropy of quantum states defined on a line. This gives rise to an interesting perspective, as it shows that to bound the decay of correlations in a suitably strong sense is sufficient for proving area laws.
 
\end{itemize}
Having said that, it should be clear that these findings can be seen as a large body of evidence that gapped many-body systems generically have the property that the 
ground state satisfies an area law.

\subsubsection{Violation of area laws}

For critical (gapless) one-dimensional models, the situation is quite distinctly different: 
The correlations decay too slowly in order to arrive at such a scaling of the entanglement entropy and the area law is violated. Still, 
the corrections 
to an area law are small. Conformal field theory suggests that  \cite{Larsen,Calabrese}
$S(\rho_A) = ({c}/{3} )\log ({l}/{a})+C$
	where $c$ is the conformal charge, $a$ the lattice spacing, and $C>0$ a non-universal constant, so in fact
\begin{equation}	
	S(\rho_A)= \Theta (\log(|A|)).
\end{equation}	
(again in Landau's asymptotic notation): It is logarithmically divergent in the size of the subsystem. 
For free fermionic models in 1D, and also for the XX model \cite{Its,FranciniLong} and the non-isotropic instance, the XY model \cite{Single},
 the precise scaling of the entanglement entropy has been rigorously proven using the {\it Fisher-Hartwig machinery} to compute {\it Toeplitz
 determinants}, again finding a logarithmic divergence, confirming a scaling that has been numerically determined in Ref.\ \cite{Latorre1}.
 For a review, again see Ref.\ \cite{Area}; for an excellent 
 non-technical introduction specifically in entanglement entropies
in quantum field theories, see Ref. \cite{CardyReview}. 
 
How do entanglement area laws in higher-dimensional {\it critical models} scale? This question is still largely open. It is known that critical free-fermionic systems violate
an area law: For a cubic lattice in ${\cal D}$ dimensions, one has
\begin{equation}
	S(\rho_A)= \Theta (L^{{\cal D}-1} \log L),
\end{equation}
which is (slightly) more than an area law would suggest \cite{AreaWolf,Klich,Halfspace}. Critical bosons, in contrast, can well satisfy an area law, even if critical
\cite{Halfspace}.

\subsubsection{Other concepts quantifying entanglement and correlations}

The entanglement entropy is a unique measure of entanglement for pure states according to some axiomatic characterisation, but this does not mean that there are not
a number of other quantifiers that  meaningfully capture the entanglement content. Importantly in the context discussed here, one may
replace the von Neumann entropy
$S(\rho_A)$ by other {\it Renyi-entropies}
\begin{equation}
	S_\alpha(\rho_A) =\frac{1}{1-\alpha}\log_2(\rho_A^\alpha),
\end{equation}
for $\alpha\geq 0$. For $\alpha\downarrow 1$, this quantity reduces to the von-Neumann entropy of $\rho_A$. One also has $S_\infty(\rho_A) = -\log_2\|\rho_A\|$,
where $\|\rho_A\|$ is the operator norm of $\rho_A$ and $S_0(\rho_A) =  \log_2 \text{rank} (\rho_A)$. These slightly more general entanglement entropies
will become important for our understanding of grasping ground states of one-dimensional models in terms of entanglement entropies. Again, for any of the above
proven instances, it is also true that a more general Renyi-entropic area laws holds true as well.

For mixed states, such as for degenerate ground states or thermal states, the entropy of entanglement is no longer physically meaningful to quantify
the degree of entanglement. 
There are essentially two strategies one can still
pursue: On the one hand, one can look at measures capturing all correlations, not only quantum correlations or entanglement. The most accepted measure quantifying correlations is
the {\it quantum mutual information}, defined as\index{Quantum mutual information}
\begin{equation}
	I(A,B):= S(\rho_A)+S(\rho_B)-S(\rho).
\end{equation}
For every Gibbs state $e^{-\beta H}/\tr(e^{-\beta H})$  of a local Hamiltonian $H$ for some inverse temperature $\beta>0$, 
this quantity is known to again satisfy an area law \cite{Correlations}, albeit with a prefactor that grows linearly in the inverse
temperature. These bounds can also in the inverse temperature 
be exponentially tightened for specific models such as the 
XX model \cite{ThermalAreaLaw}. \index{Gibbs state}
On the other hand, mixed state entanglement measures can be employed that still capture genuinely quantum correlations even in mixed-state situations when
quantum and classical correlations are intertwined. One of the most prominent such entanglement measure 
is the so-called {\it negativity} \cite{PhD,VidalNegativity,PlenioNegativity}, 
defined as 
\begin{equation}
	E (\rho) = 
\| \rho^{T_A}\|_1-1, 
\end{equation}
where $\|.\|_1$ is the trace norm ($\|O\|_1 = \tr(|O|)$ for an operator $O$) and $\rho^{T_A}$ is the partial transpose of $\rho$, so the operator that is 
obtained by taking a partial transpose with respect to the tensor factor belonging to subsystem $A$. {\it Entanglement negativities} have been studied in several contexts
\cite{Harmonic,CalabreseNegativity,CastelnovoNegativity}.
Since the logarithmic version, called {\it logarithmic negativity} $\log_2 \| \rho^{T_A}\|_1$ is an upper bound to the entanglement entropy, such quantities have also extensively
been used in the past to derive area laws for entanglement quantifiers, 
even for non-degenerate ground states when entanglement entropies are no longer applicable.\index{Entanglement negativity}

\subsubsection{Entanglement spectra}

The entropy of entanglement is just a single number, but it should be rather obvious that more detailed information is revealed when the entire spectrum of $\rho_A$ is considered.
In fact, the collection of all Renyi entropies of $\rho_A$ gives precisely the same information as the spectrum of $\rho_A$ itself. Given a state $\rho_A$, it is meaningful
to consider the {\it entanglement Hamiltonian} $H_A$ for which $\rho_A= e^{-H_A}$. In fact, the full entanglement spectrum (so the spectrum of $\rho_A$) reveals a lot
of important information about the quantum many-body system and is receiving a significant attention in the context of {\it topological systems} and
{\it boundary theories} \cite{PeschelEntanglementDMRG,Spectrum,Spectrum3,Boundary}.\index{Entanglement spectrum}


\subsection{The notion of the `physical corner of Hilbert space'}

\subsubsection{The exponentially large Hilbert space}
We have seen that ground states of gapped quantum many-body models exhibit little entanglement, in fact much less that they could feature. 
Interactions are short ranged, which not only
means that correlations decay strongly, but also that there is very little entanglement present in the above sense.
This is a basic, but at the same time very important, observation: It appears in this
light that natural ground states (and Gibbs states) seem to explore only a very tiny fraction of Hilbert respectively state space that would in principle be
available. Let us not forget that Hilbert space is exceedingly big: For a spin
system of size $n$ and local dimension $d$, the dimension scales as
\begin{equation}
	\dim ({\cal H}) = O(d^n).
\end{equation}
It should be clear that already for moderately sized systems, state vectors can no longer be stored on a computer in order to numerically solve the system in exact diagonalisations
(naively requiring $O(d^{3n})$ operations). Surely,
one can and must 
heavily exploit symmetries and sparsity patterns in the Hamiltonian to reduce the effective subspace that has to be considered, and then for moderately sized systems,
exact diagonalisations can provide impressive results \cite{Exact}. In any case, needless to say, one will encounter a scaling of the 
dimension of the relevant subspace that is exponential in the system size.

\subsubsection{Small subset occupied by natural states of quantum many-body models}

The key insight here is that that the pure state exhibiting low entanglement in the sense of satisfying an area law constitute a very 
small subset of all pure states. What is more, this subset can be well approximated by tensor network states. In the end, the reason for tensor network methods 
to provide such powerful tools is rooted in the fact that natural ground states satisfy area laws (or show small violations thereof). 
In this sense, one might well say that the exponentially large Hilbert space
`is a fiction', and the `natural corner of Hilbert space' constitutes only an exceedingly small fraction thereof. This somewhat wordy notion will be made more precise in a minute.
One should not forget, after all, that not only ground states occupy a tiny fraction of Hilbert space, but the same holds true for all efficiently preparable quantum states:
Not even a {\it quantum computer}
could efficiently prepare all states \cite{Illusion}, in fact a far cry from that. (The picture depicts not only the subset of 
ground states of local Hamiltonians, but also matrix product states of some bond dimension $D$.)

\begin{figure}[h!]
 \centering
\includegraphics[width=0.85\textwidth]{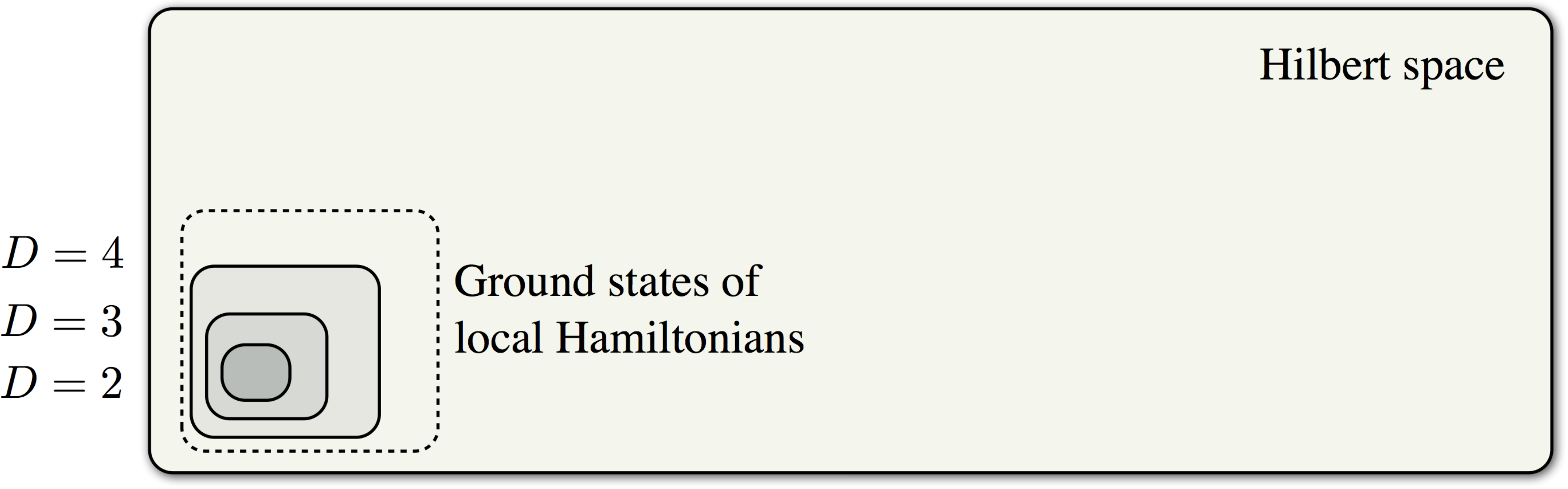}
\end{figure}

\section{Matrix product states}

We now turn to exploiting this insight when grasping quantum many-body states in terms of tensor network states. We start by introducing a commonly used and quite
illustrative graphical notation. We will
then discuss in great detail the concept of a matrix product state which features in the highly successful density-matrix renormalisation group
(DMRG) algorithm \cite{White,Scholl,SchollwoeckAge,Rommer,Periodic}. It is worth noting that the history of this concept is actually quite remarkable: 
It actually appeared several times independently in the literature. {\it Finitely correlated states} as formulated in an algebraic picture in the language of mathematical
physics \cite{FCS} can be 
viewed as translationally invariant infinite matrix product states. In the same year as finitely correlated states were proposed, they independently emerged 
implicitly in the seminal work on DMRG by Steve White \cite{White} in the context of condensed-matter physics -- even if it took until much later until the close
connection was spotted \cite{Rommer,Kluemper}. 
In the meantime, the DMRG method is routinely explained in terms of matrix product states \cite{SchollwoeckAge}, a mindset that 
we will also follow. Yet independently, the concept of a {\it tensor train
decomposition} \cite{Hackbusch}
emerged in the mathematics literature, which was again found to be essentially equivalent to the concept of a matrix product state.

\subsection{Preliminaries}

 A {\it tensor} can be represented as a multi-dimensional array of complex numbers. The dimensionality of the array required to represent it is called the {\it order} of the tensor.
 A scalar can be viewed as a tensor or order $0$, a vector is seen as a tensor of order $1$, a matrix would be tensor of order $2$, and so on. We will make extensive use of the
 graphical notation that goes back to Penrose to represent tensors:  We will graphically represent tensors
 as boxes, having a number of edges defined by the order of the tensor. 
 This is then how a scalar looks like,
 \begin{figure}[h!]
 \centering
\includegraphics[width=0.06\textwidth]{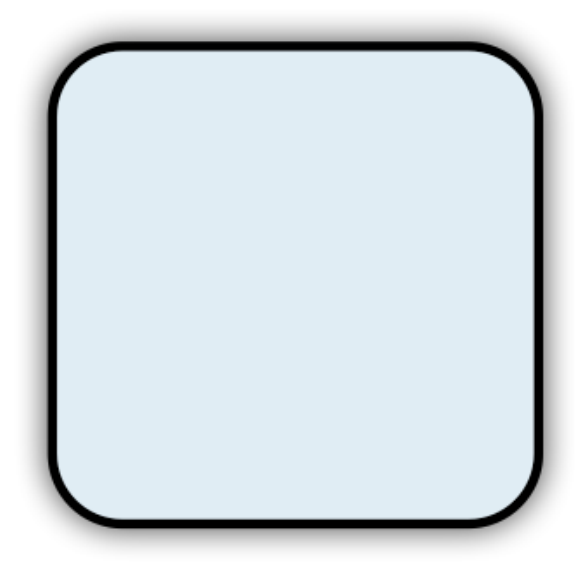}
\end{figure}

these are vectors and dual vectors,
 \begin{figure}[h!]
 \centering
\includegraphics[width=0.25\textwidth]{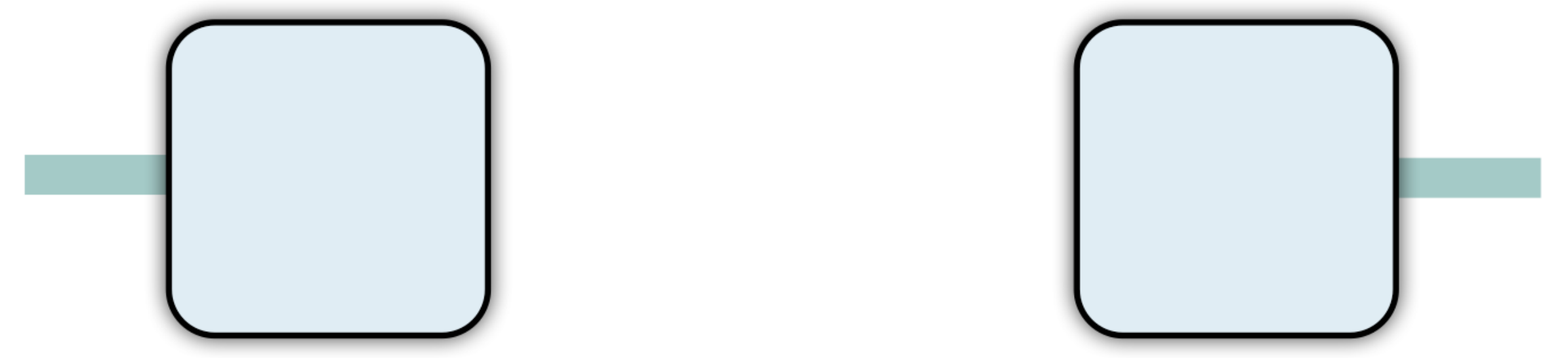}
\end{figure}
 
 and this
  \begin{figure}[h!]
 \centering
\includegraphics[width=0.109\textwidth]{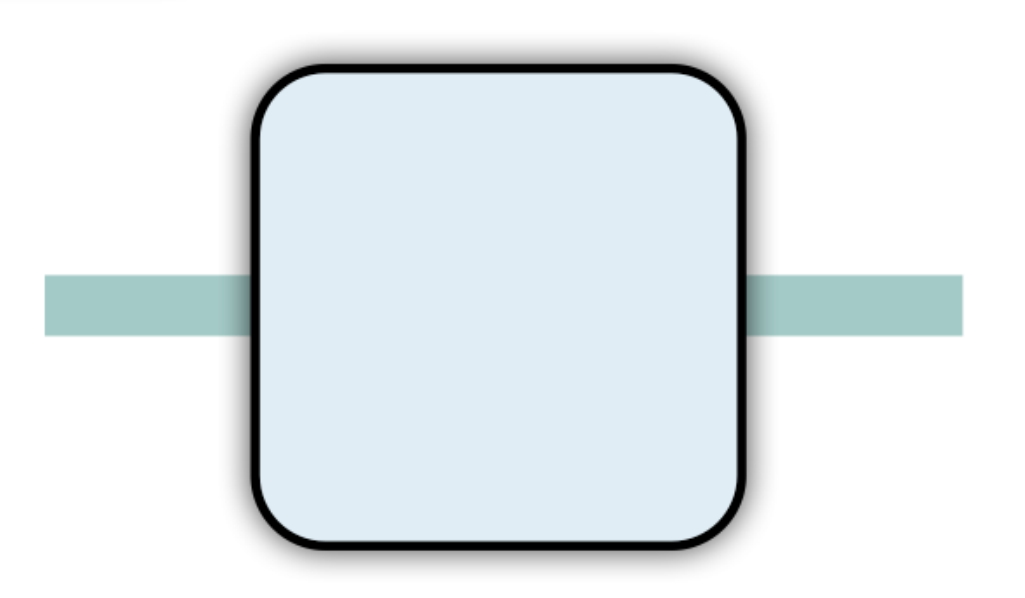}
\end{figure}

 corresponds to a matrix.  A contraction of an index amounts to summing over all possible values an index takes corresponding to shared edge. For example, a matrix product
 $A=BC$ of matrices $A,B,C\in \cc^{N\times N}$ amounts to\index{Tensor network}
 \begin{equation}
 	C_{\alpha,\beta}= \sum_{\gamma=1}^N A_{\alpha,\gamma} B_{\gamma,\beta},
 \end{equation}
 so here the common index $\gamma$ is being contracted. 
Again, we can graphically represent this as
 \begin{figure}[h!]
 \centering
\includegraphics[width=0.32\textwidth]{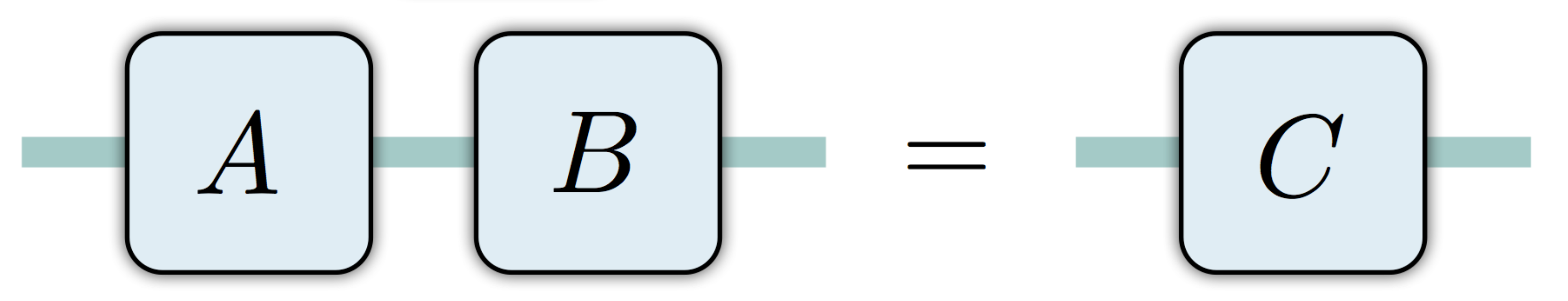}.
\end{figure}

 The trace is a contraction of two indices of the same tensor, graphically
  \begin{figure}[h!]
 \centering
\includegraphics[width=0.125\textwidth]{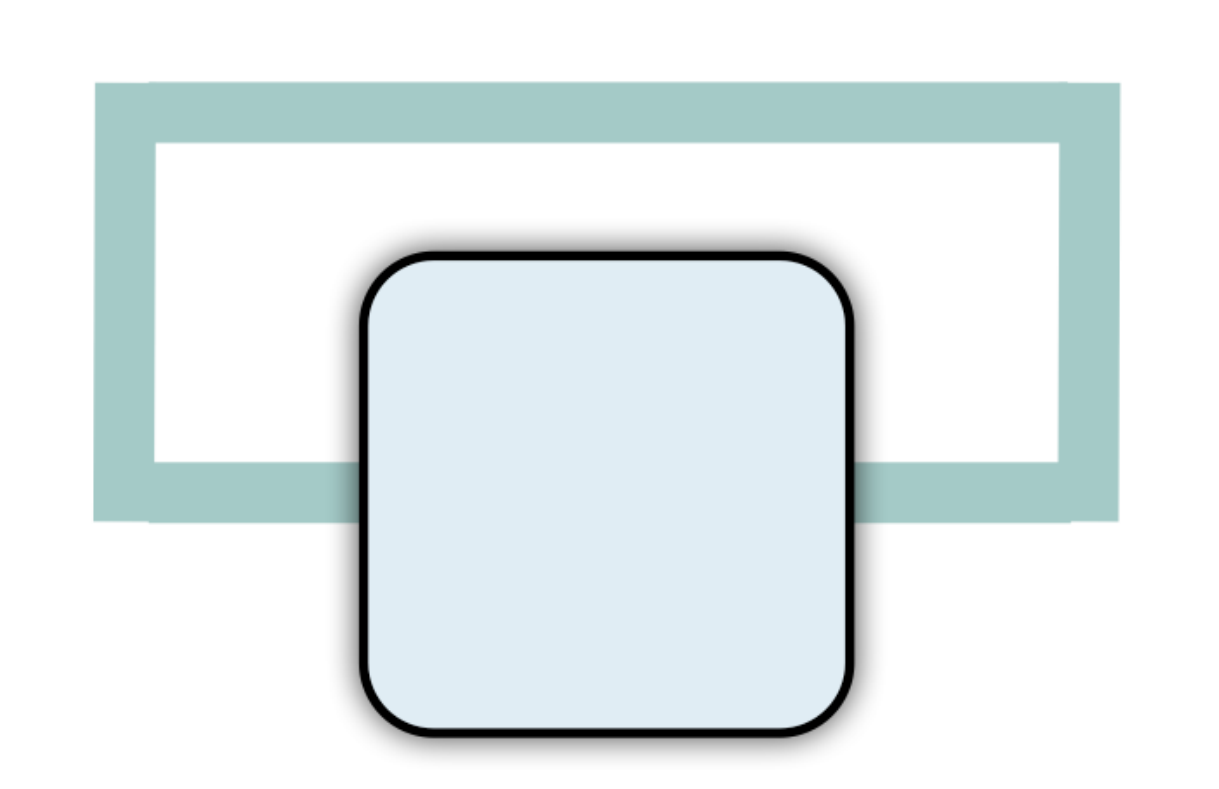},
\end{figure}

and a partial trace is 
  \begin{figure}[h!]
 \centering
\includegraphics[width=0.115\textwidth]{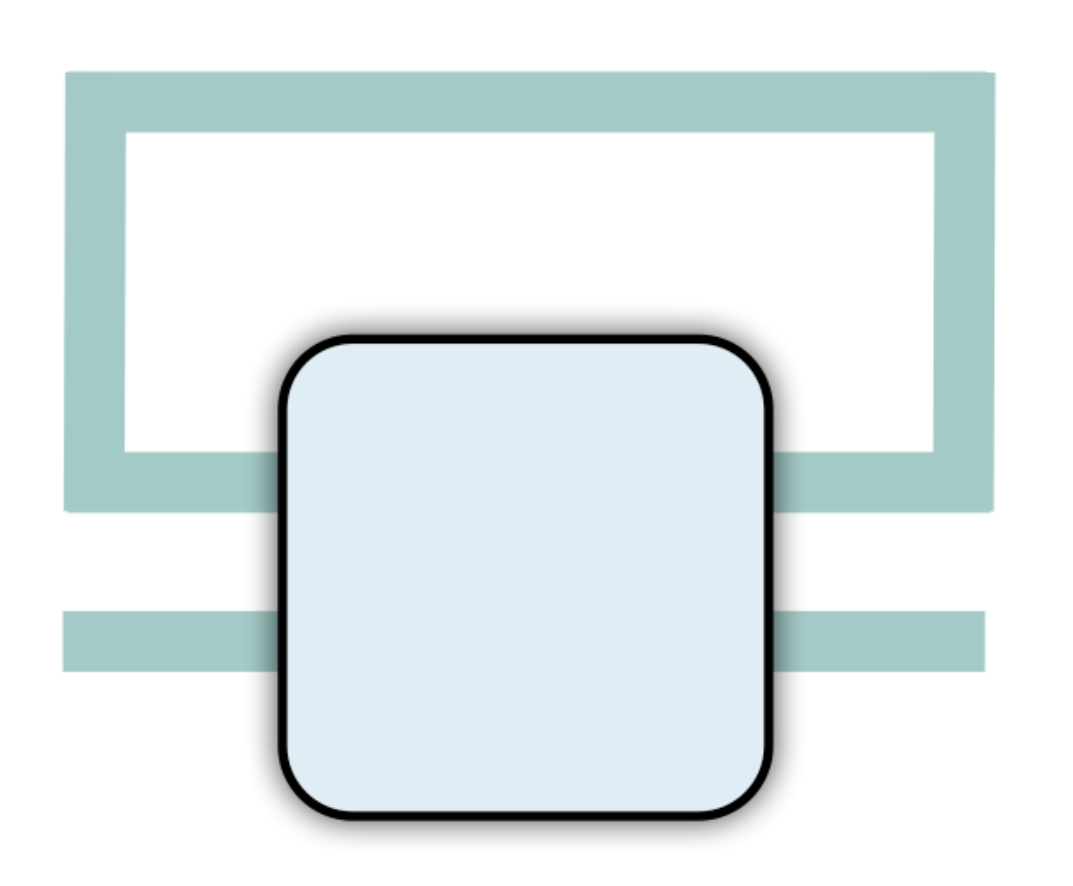}.
\end{figure}

 A scalar product 
   \begin{figure}[h!]
 \centering
\includegraphics[width=0.165\textwidth]{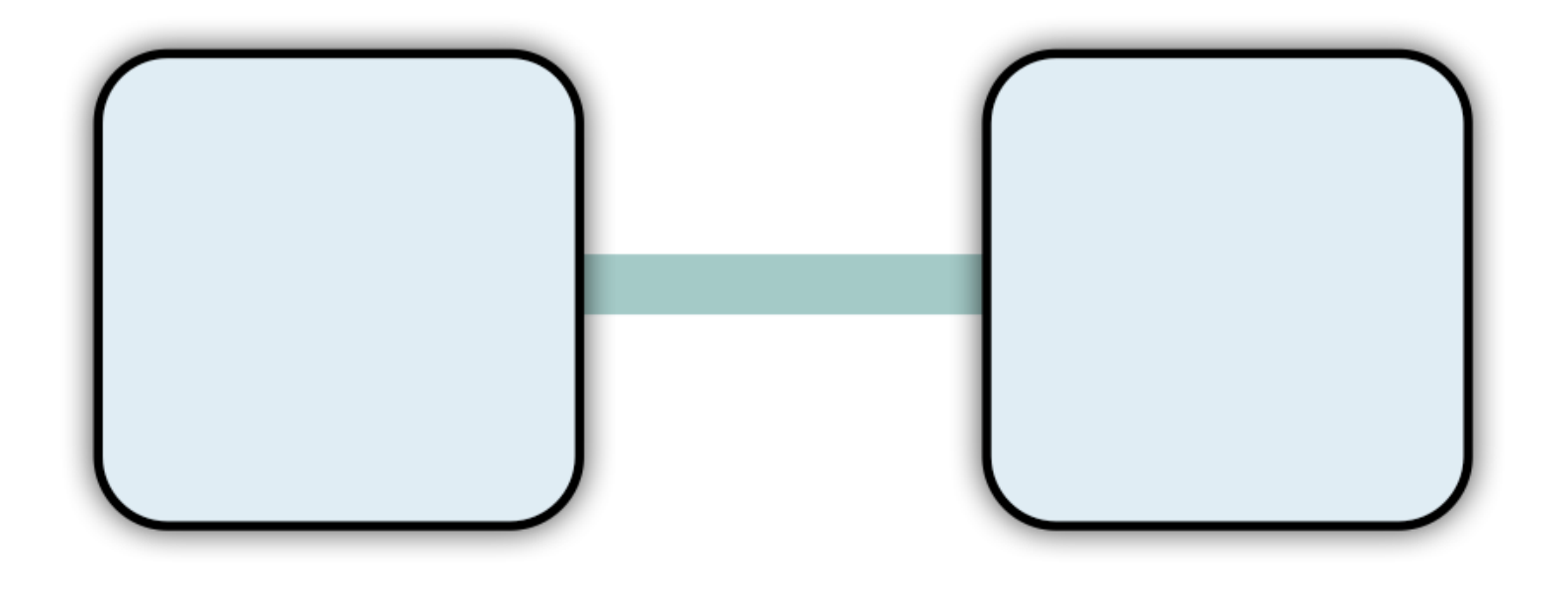}
\end{figure}

 looks like this.
 The beauty and ease of this picture should be rather obvious.\footnote{We will take a pragmatic viewpoint here and will swipe some mathematical fine print concerning such
 graphical tensor network representations of tensors over complex vector spaces under the carpet which we should not too much worry about, however.}
 An index that is not contracted is naturally called an {\it open index}.
A {\it contraction of a tensor network}, like this one,
   \begin{figure}[h!]
 \centering
\includegraphics[width=0.35\textwidth]{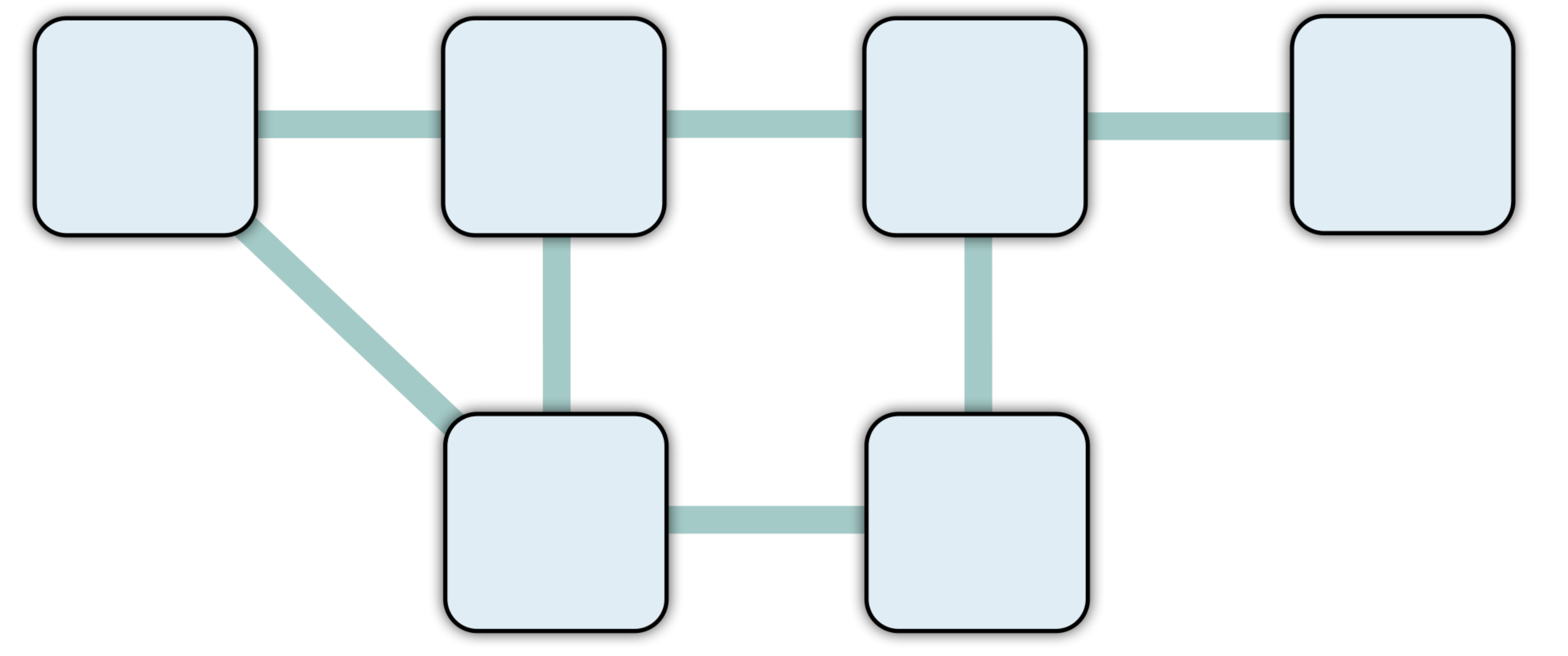}
\end{figure}

amounts to contracting all indices that are not open. Where would one best start with such a contraction? Indeed,
the order very much matters as far as the complexity of the problem is concerned. The scaling of the effort of contraction in the dimension of the involved
tensors is highly dependent on the contraction order, and to find the optimal order of pairwise contractions is a computationally hard problem in its own right. 
In practice, one often finds good contraction orders by inspection, however.

We now turn to the graphical representation of what we are mainly interested in, namely state vectors of 
quantum many-body spin systems with $n$ degrees of freedom. 
An arbitrary state vector $|\psi\rangle\in (\cc^d)^{\otimes n}$
\begin{equation}\label{sv}
	|\psi\rangle = \sum_{j_1,\dots, j_n=1}^d c_{j_1,\dots, j_n} |j_1,\dots, j_n\rangle = \sum_{j_1,\dots, j_n=1}^d c_{j_1,\dots, j_n} |j_1\rangle\otimes \dots\otimes | j_n\rangle
\end{equation}
with coefficients $c_{j_1,\dots, j_n}\in \cc$ for all indices can be represented by
  \begin{figure}[h!]
 \centering
\includegraphics[width=0.48\textwidth]{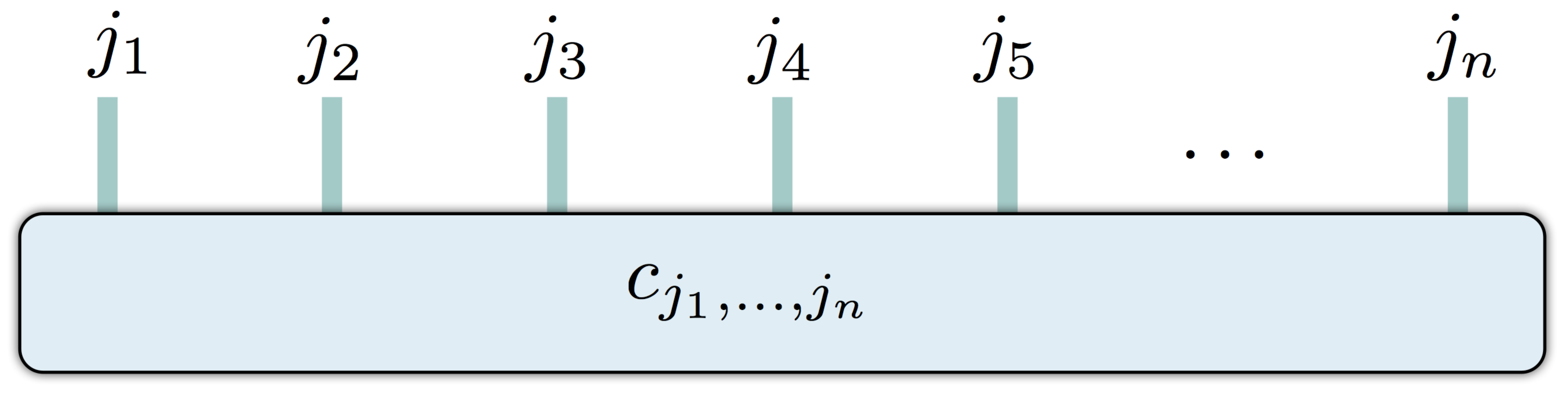}
\end{figure}
so by a box with $n$ edges, sometimes also called `physical edges' for obvious reasons.

\subsection{Definitions and preparations of matrix product states}

\subsubsection{Definition for periodic boundary conditions}

The definition of matrix product states takes the above tensor and `breaks it down' to smaller components that are being contracted. A {\it matrix product state} 
\index{Matrix product state}
\cite{FCS,MPSSurvey}
of `bond dimension' $D$ (with periodic boundary conditions) 
is a pure state with a state vector of the form
\begin{equation}\label{mps}
	c_{j_1,\dots, j_n} = \sum_{\alpha,\beta,\dots,\omega=1}^D A^{(1)}_{\alpha,\beta;j_1} A_{\beta,\gamma; i_2}^{(2)}\dots A_{\omega,\alpha;j_n}^{(n)}= 
	\tr(
	A_{j_1}^{(1)} A_{j_2}^{(2)}\dots A_{j_n}^{(n)}
	),
\end{equation}
where the trace and the matrix product are taken over the contracted indices, leaving the physical indices $j_1,\dots, j_n$ open. For a fixed
collection of physical indices, the coefficients are hence obtained by considering matrix products of matrices, hence `matrix product state'.
In a graphical notation, this can be represented as
  \begin{figure}[h!]
 \centering
\includegraphics[width=0.47\textwidth]{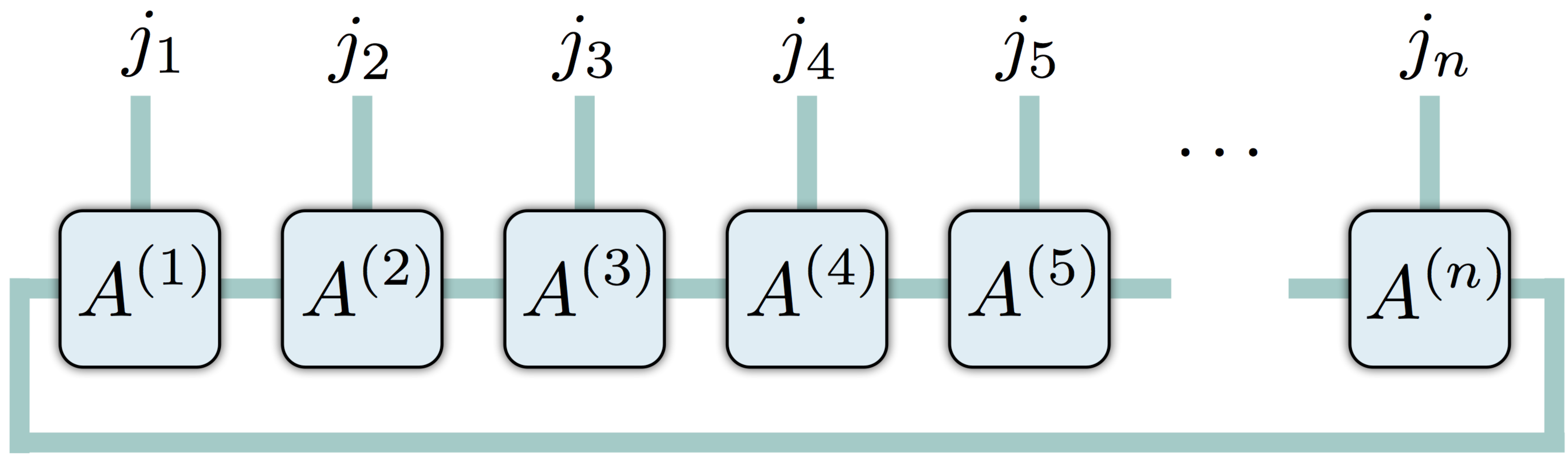}.
\end{figure}

That is to say, each individual tensor is represented as
  \begin{figure}[h!]
 \centering
\includegraphics[width=0.18\textwidth]{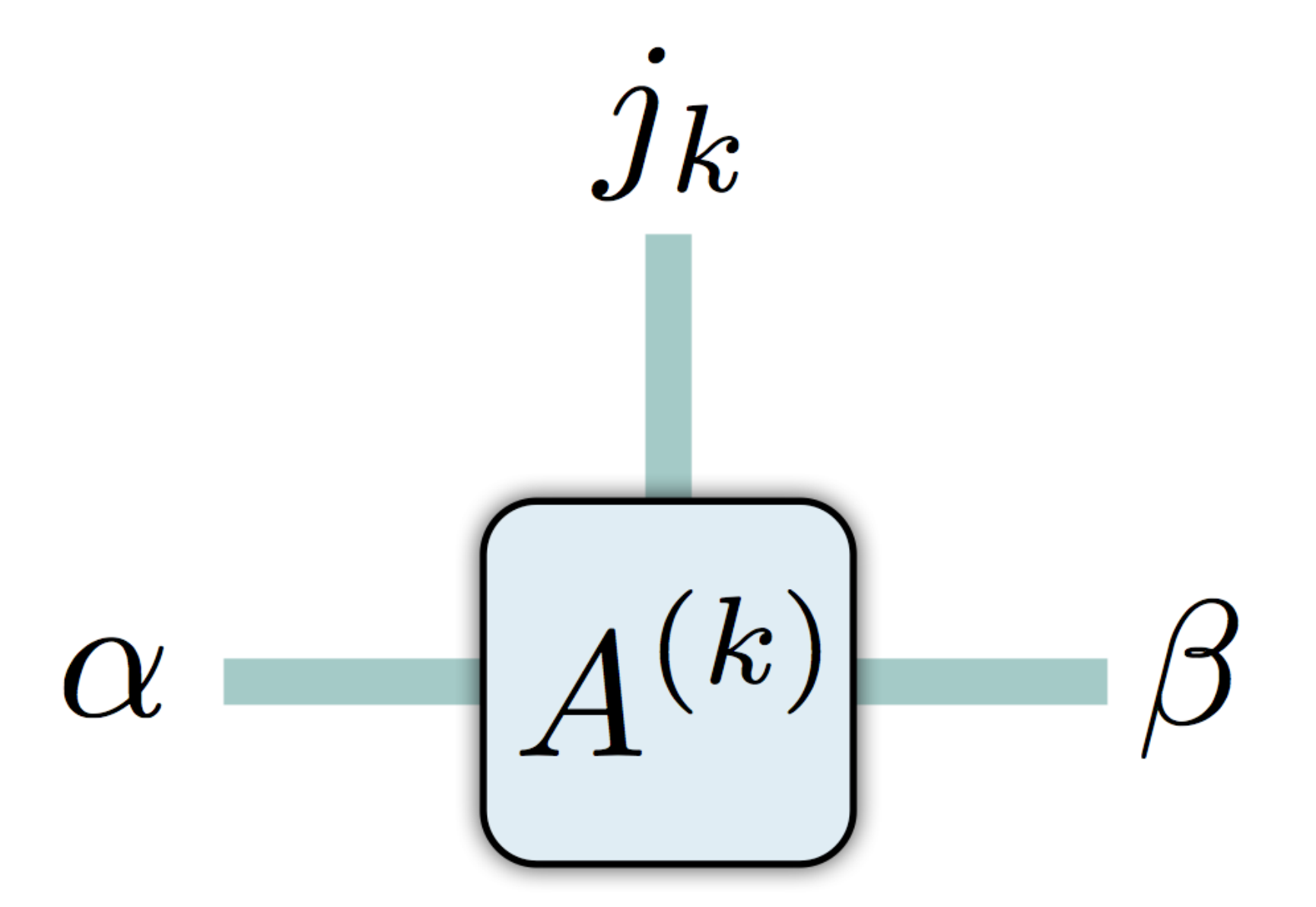}
\end{figure}

 and via contraction, one arrives at the above expression. The line connecting the end tensors reflects the trace in the above expression.
 This graphical notation will remain very handy in what follows.

 So what is $D$, the bond dimension? As such, it does not have a direct physical correspondence; 
 this parameter can be viewed as a `refinement parameter'. It will also soon
 become clear why it is called a bond dimension and we will turn to its significance in a minute.
 Matrix product states constitute the in many ways most important instance of a tensor network state. They are
 of key importance both in analytical approaches as well as in numerical ones, most prominently in the density-matrix renormalisation group approach. Since we will frequently
 refer to such states, we will from now on commonly abbreviate them as MPS.
   
 \subsubsection{Variational parameters of a matrix product state}
 
 We note a first important property of a matrix product state: It is described by very few numbers. While a general state vector of a system composed of $n$ spin-$d$ systems
 is defined by $O(d^n)$ many real parameters, an MPS of bond dimension $D$ can be represented by $O(n d D^2)$ many real parameters. For constant $D$, this is linear in $n$,
 as opposed to exponential in $n$: so this ansatz 
 gives rise to a drastic reduction of the number of variational parameters, to say the least. At the same time it is true that $D$ can be 
 taken large enough that every state vector of a finite system
 can be represented as an MPS, if one allows $D$ to grow exponentially in $n$ as well. Yet, this is actually not the main point of the definition of
 a matrix product state. 
 
 The key insight -- one that should become increasingly clear -- is that already for small bond dimension $D$, an MPS is an extraordinarily good
 approximation of natural states emerging in physical systems. The larger the bond dimension, so the `refinement parameter' $D$, the larger is the set of states that can be
 represented, and hence usually the quality of the approximation of natural states. If one takes $D=1$, then the above matrices merely become complex numbers and
 one obtains a product state, in a variational set that is sometimes referred to
 as a {\it Gutzwiller} variational state, a variant of a {\it mean-field approach}.

\subsubsection{Matrix product states with open boundary conditions}

The above expression corresponds to matrix product states for periodic boundary conditions. For open boundary conditions, 
the matrix $A^{(1)}$ is taken to be no longer a matrix from $\cc^{D\times D}$, but  $A^{(1)}\in \cc^{1\times D}$ so as a row vector. Similarly
$A^{(n)}\in \cc^{D\times 1}$ so it is a column vector. Then the expression becomes 
\begin{equation}\label{omps}
	c_{j_1,\dots, j_n} = \sum_{\alpha,\dots,\omega=1}^D A^{(1)}_{\alpha;j_1} A_{\beta,\gamma; i_2}^{(2)}\dots A_{\omega;j_n}^{(n)}= 
	A_{j_1}^{(1)} A_{j_2}^{(2)}\dots A_{j_n}^{(n)},
\end{equation}
and the graphical expression
  \begin{figure}[h!]
 \centering
\includegraphics[width=0.47\textwidth]{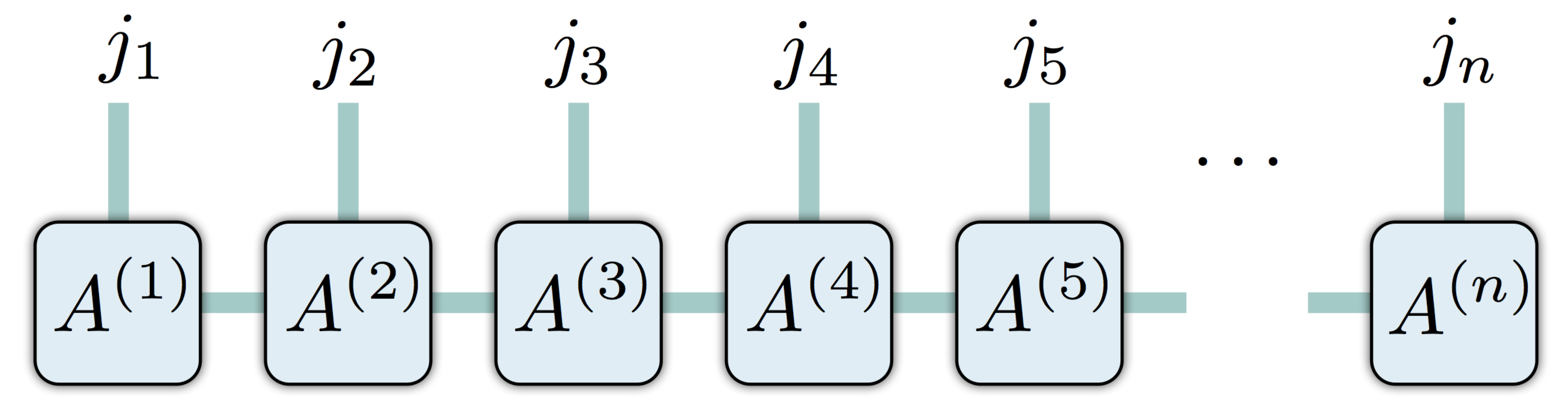}.
\end{figure}

\subsubsection{Area laws and approximation with matrix product states}

What is the significance of area laws in the context of matrix product states?
It is easy to see that for any subset $A$ of consecutive sites of the lattice $S(\rho_A)=O(\log (D))$ for a matrix product state, so the entanglement entropy is bounded from above
by a constant in $n$. That is to say, MPS satisfy an area law. The behaviour of the entanglement scaling 
is therefore the same for matrix product states as for ground states of gapped models.
But indeed, an even stronger converse statement is true: Every state that satisfies an area law can be efficiently approximated by a matrix product state.

There is a bit of fine-print associated with this statement:
On the one hand, the precise wording of this statement makes use of Renyi entropic entanglement entropies as discussed above, and not the more standard 
entanglement entropies based on the von-Neumann entropy. 
(Yet, for all situations where a detailed understanding has been reached, this does not make a difference anyway, so we can be rather relaxed about it.)
On the other hand, any such statement is surely one about a family of states of lattice systems of increasing system size, rather than for a single
state. So precisely,
if for a family of state vectors $|\psi_n\rangle$ 
there exist constants $c,C>0$ such that for all $0<\alpha<1$ the Renyi entropies of the reduced state of any subsystem $A$ of the one-dimensional system satisfy
\begin{equation}
	S_\alpha(\rho_A) \leq c \log(n) + C,
\end{equation}
then it can be efficiently approximated by an MPS (so the bond dimension will have to grow polynomially with $n$, the system size, and $1/\varepsilon$, where 
$\varepsilon>0$ is the approximation error).

\subsubsection{Preparation from maximally entangled pairs}

There are two alternative convenient descriptions of MPS. The first one originates from a hypothetical preparation from maximally entangled states. 
This prescription explains why MPS can also be captured as `projected entangled pair states' for one-dimensional spin chains. Let us assume, for that purpose,
that each site of a one-dimensional spin chain is not composed of a single $d$-dimensional system, but in fact of two `virtual systems' of dimension $D$. Each one
of the pair of virtual systems is {\it maximally entangled} with the respective
neighbouring system, which means that the state of this pair is described by the state vector
\begin{equation}\label{maxent}
	|\omega\rangle = \sum_{j=1}^d |j,j\rangle.
\end{equation}
Such a pure state is indeed `maximally entangled' in that the entanglement entropy of each subsystem takes the maximum value $\log_2(D)$. These entangled states
are referred to as `bonds', further motivating the above term of a `bond dimension'. On this system, we apply linear maps 
$P^{(j)}:\cc^D\otimes \cc^D\rightarrow \cc^d$ for each of the sites
$j=1,\dots, n$, projecting two $D$-dimensional systems to a single system of the physical dimension $d$. We can hence capture the state vector as
\begin{equation}
	|\psi\rangle = (P^{(1)}\otimes \dots \otimes P^{(n)}) |\omega\rangle^{\otimes (n-1)}.
\end{equation}
\index{Projected entangled pair state}
This prescription, even though it may look somewhat baroque at this point, surely defines
a valid state vector in $(\cc^d)^{\otimes n}$. The claim now is that this is an alternative
description of a matrix product state. How can that be?
\newpage
Let us write each of the the linear maps as
\begin{equation}
	P^{(j)}  =\sum_{k=1}^d \sum_{\alpha,\beta=1}^D A_{\alpha,\beta;k}^{(j)} |k\rangle\langle\alpha,\beta|
\end{equation}
for $j=1,\dots, n$, for periodic boundary conditions,
graphically
\begin{figure}[h!]
 \centering
\includegraphics[width=0.53\textwidth]{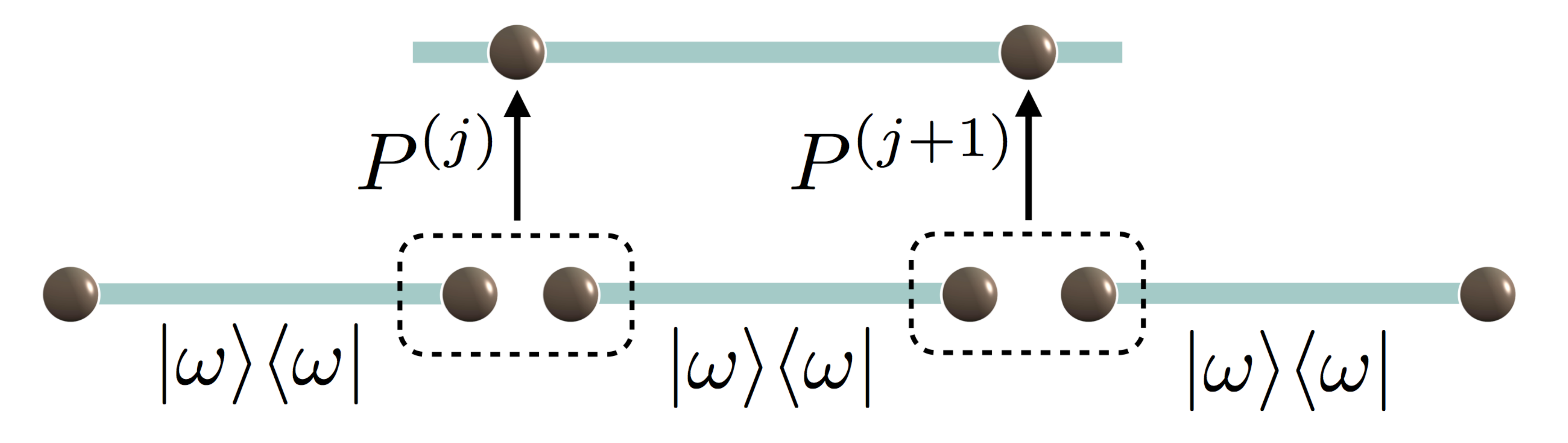}.
\end{figure}

Then, a moment thought reveals that we again end up in a state vector with coefficients precisely of the form of Eq.\ (\ref{mps}). So the matrices that define the linear projection reappear in a different role in the definition of the matrix product state. An interesting side remark is that in this picture it is also particularly clear that MPS satisfy area laws, with $\log_2(D)$ being the maximum value the entanglement entropy can take. This picture will be particularly intuitive when generalising the idea of MPS to higher dimensional physical systems.

\subsubsection{Sequential preparation picture}

The second alternative convenient description of MPS relates to a sequential preparation of quantum states, an idea that was implicitly already present in Ref.\ \cite{FCS}. 
Here, one starts off in a spin chain of local dimension $d$
prepared in $|0\rangle^{\otimes n}$ and 
lets a quantum system of dimension $D$ sequentially interact with each of the $n$ constituents. At the end, one makes sure that the system is disentangled. It turns out that the state vectors generated in this way are exactly the MPS with open boundary conditions (see, e.g., Ref.\ \cite{MPSSurvey}). More specifically, let 
$\sum_{j=1}^d\sum_{\alpha,\beta=1}^D A_{\alpha,\beta;j}^{(k)}|\alpha,j\rangle\langle \beta,0|$ be an operation on $\cc^D\otimes \cc^d$ with
\begin{equation}
	\sum_{j=1}^d (A^{(k)})^\dagger_j A_j^{(k)}=\id
\end{equation}
(we will see below that this can always be chosen to be true) for each $k$, 
then one obtains an MPS with open boundary conditions of the form
as in Eq.\ (\ref{omps}). This construction is interesting in many ways: To start with, this procedure gives rise to an efficient preparation of MPS, and there are several
physical systems where one can readily think of systems sequentially interacting in this way (for example for atoms passing through cavities). 
In fact, in a slightly different language, MPS are discussed in the 
context of {\it quantum memory channels}, where the memory is encoded in the $D$-dimensional system passed from one site to the other. The second insight is that this
sequential interaction picture plausibly motivates the exponential decay of correlation functions that we will learn about soon: All quantum correlations have to be mediated to the 
$D$-dimensional `ancilla' system that is sequentially interacting.

\subsubsection{Translationally invariant matrix product states}

In the above definition, we have taken all matrices to be different. Of course, we can also in a translationally invariant ansatz choose
them to be all the same, so take
for periodic boundary conditions
\begin{equation}
	A_{\alpha,\beta;k}^{(j)} = A_{\alpha,\beta;k}
\end{equation}
for all $\alpha,\beta=1,\dots, D$, all $k=1,\dots, d$ and all sites $j=1,\dots, n$. 

\begin{figure}[h!]
 \centering
\includegraphics[width=0.46\textwidth]{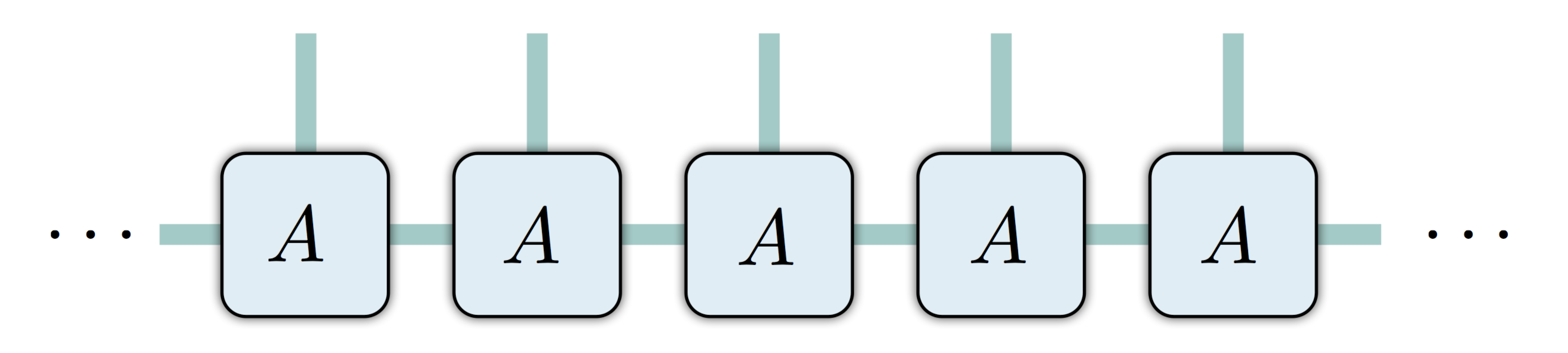}.
\end{figure}

Such translationally invariant MPS make a lot of sense in analytical considerations,
and obviously capture translationally invariant models well. They are specifically important when considering the thermodynamic limit $n\rightarrow \infty$.
In numerical considerations, it is for good reasons
often advisable to break the symmetry and use different matrices per site, 
even if the Hamiltonian as such is
translationally invariant.

\subsubsection{Successive Schmidt decompositions}

The canonical form of an MPS can also be reached by making use of a {\it successive Schmidt decomposition}. 
This was first highlighted in Ref.\
\cite{GVidal}. We will be brief here, but explain the basic 
idea: Generally, a Schmidt decomposition of a state vector $|\psi\rangle\in \cc^{d_A}\otimes \cc^{d_B}$ 
of a system consisting of two parts $A$ and $B$
can be written as 
\begin{equation}
	|\psi\rangle = \sum_{j=1}^{\min(d_A,d_B)} \lambda_j |\psi^{(A)}_j\rangle \otimes |\psi^{(B)}_j\rangle ,
\end{equation}
with suitable orthonormal bases $\{|\psi_{j}^{(A)}\rangle\}$ and $\{|\psi_{j}^{(B)}\rangle\}$ of the respective Hilbert spaces, called left and right Schmidt vectors and $\lambda_j\geq 0$
for all $j$. Why is this remarkable? Because there is only a single sum, not
a double sum. One can indeed now arrive at the canonical form in a one-dimensional MPS by starting from the left side and performing the Schmidt decomposition 
between site $\{1\}$ and the complement $\{2,\dots, n\}$ in $V$. Then one can expand the left Schmidt vectors in the original basis and continue by performing a Schmidt
decomposition of the right Schmidt vectors between $\{2\}$ and $\{3,\dots, n\}$ and so on, to arrive at the normal form.

\subsection{Computation of expectation values and numerical techniques}

\subsubsection{Computation of expectation values}

How can we compute expectation values of MPS? Of course, the entire idea of a tensor network state only makes sense if we have a handle on meaningfully (at least
approximately) computing expectation values $\langle \psi| O|\psi\rangle $ of local observables $O$. 
At this point, we have good reasons to hesitate, however, and to become a bit worried. The fact that we can describe an MPS by a few parameters alone
does {\it not} necessarily imply that we can also efficiently
compute the expectation value. For example, there are operations known, such as the permanent of a matrix,
that cannot be computed in time polynomial in the dimension of the matrix (permanent is in the complexity class $\# P$).

But let us see how far we get: Let us assume that $O$ is a local term that is supported on neighbouring 
sites $l, l+1$ only, so
\begin{equation}
	O = \sum_{j_l,j_{l+1}=1}^d
	 \sum_{k_l,k_{l+1}=1}^d
	 O_{j_l,j_{l+1};k_l,k_{l+1}}
	 |j_l,j_{l+1}\rangle\langle k_l,k_{l+1}| .
\end{equation}
We suppress an index specifying the support.
It should be clear that the same strategy can be applied to local
terms with larger locality regions, so let us stick to nearest neighbour interaction terms for simplicity of notation.
 We pick open boundary conditions (but not necessarily a translationally invariant ansatz) to render the discussion more transparent.
We now start from
\begin{eqnarray}
	\langle\psi| O |\psi\rangle &=& \sum_{j_1,\dots, j_n=1}^d
	\sum_{k_1,\dots, k_n=1}^d
	\bar{c}_{k_1,\dots, k_n}
	c_{j_1,\dots, j_n}
	\delta_{j_1,k_1}\dots
	\delta_{j_{l-1},k_{l-1}}\nonumber\\
	&\times&
	O_{j_l,j_{l+1}, k_l, k_{l+1}} 
	\delta_{j_{l+2},k_{l+2}}\dots
	\delta_{j_n,k_n}.
\end{eqnarray}
This expression looks quite messy.
Resorting to the graphical notation, this can be more transparently be represented as
\begin{figure}[h!]
 \centering
\includegraphics[width=0.49\textwidth]{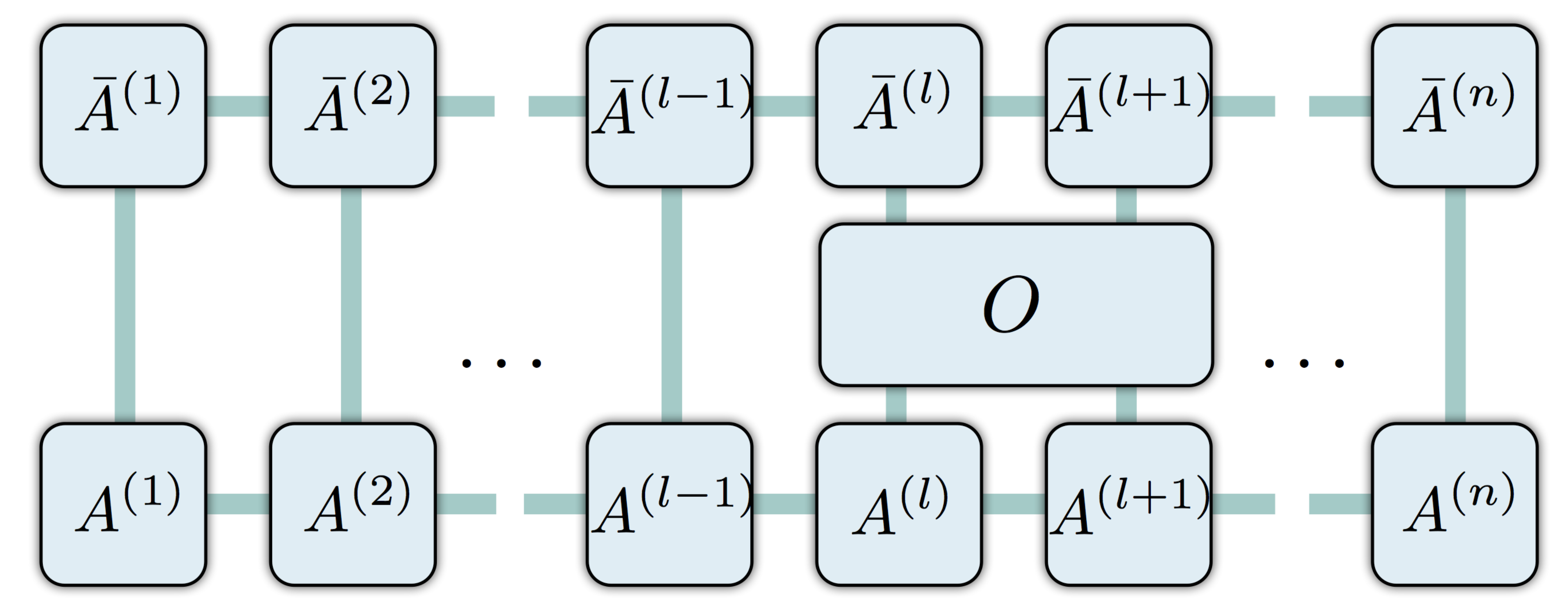}.
\end{figure}
\newpage

Naively formulated, we would clearly encounter $2n$ independent sums. In the
light of this observation, one would hence be tempted to think that one needs in $n$
exponential computational effort to compute expectation values. Then much of the
advantage of a tensor network state would disappear, needless to say.

It is one of the key insights that expectation values of local observables can nevertheless be efficiently computed -- one only needs to
contract the tensor network in a smart order. Let us remind ourselves that the contraction effort is not independent on the actual order by which the contraction is
performed. We start by investigating the left hand side of the tensor network: We can contract one index and write for the {\it left boundary condition}
\begin{equation}
	L_{\alpha,\beta}:= \sum_{j=1}^d A_{\alpha;j}^{(1)} \bar{A}_{\beta;j}^{(1)} ,
\end{equation}
graphically represented as
\begin{figure}[h!]
 \centering
\includegraphics[width=0.23\textwidth]{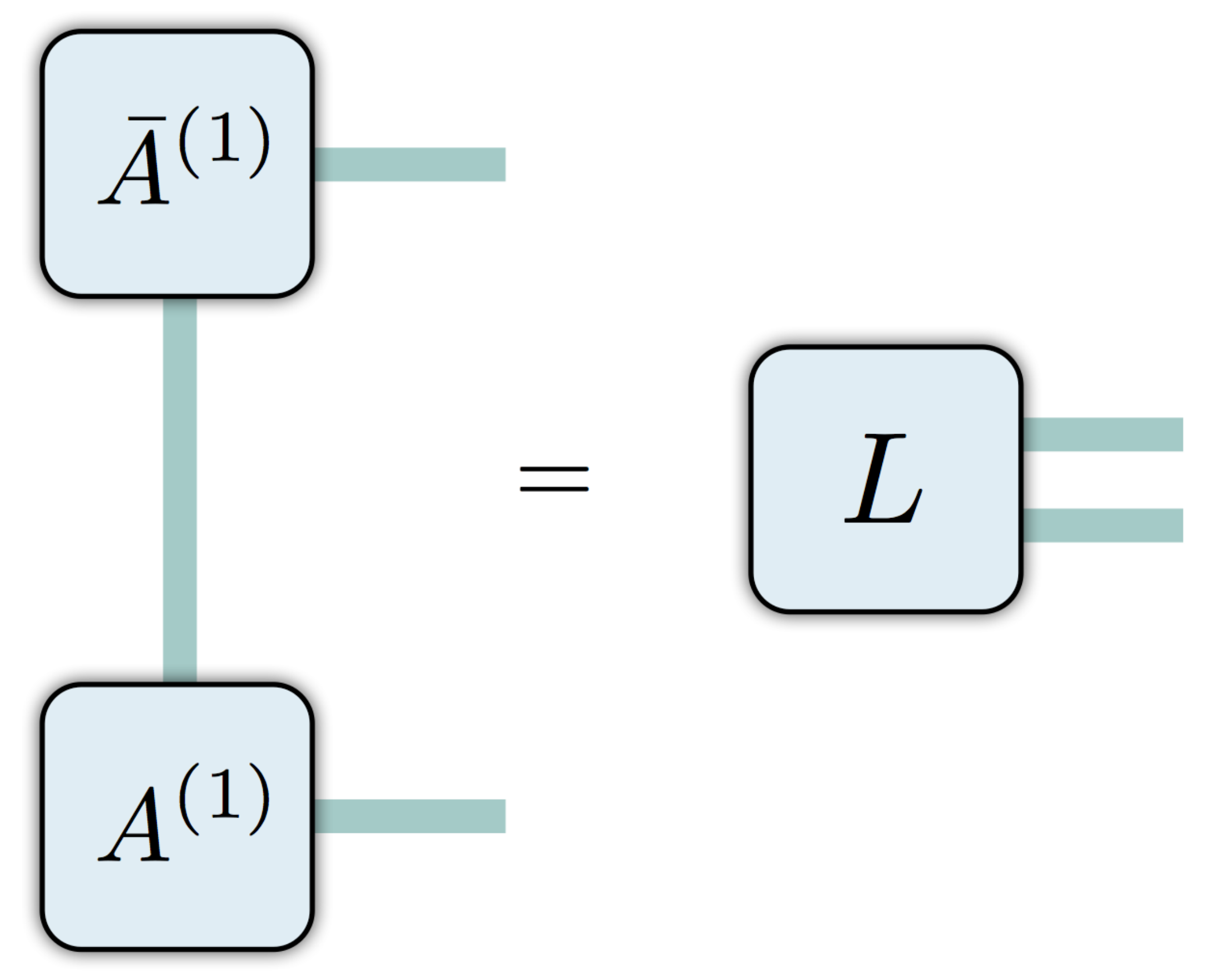}.
\end{figure}
\index{Transfer operator}

Now this again gives rise to a vector from $\cc^{1\times D^2}$: We have `doubled' the indices and hence encounter an edge associated with dimension $D^2$ instead of $D$.
We can now proceed as before, again contracting physical indices. In this way, we arrive at the {\it transfer operator} $E_{\id}^{(k)}$. This operator has 
components
\begin{equation}
	(E^{(k)}_\id)_{\alpha,\beta;\gamma,\delta} = \sum_{j=1}^d A_{\alpha,\beta;j}^{(k)} \bar{A}_{\gamma,\delta;j}^{(k)}.
\end{equation}
\newpage
At this point, the graphical notation seems
straightforward,

\begin{figure}[h!]
 \centering
\includegraphics[width=0.29\textwidth]{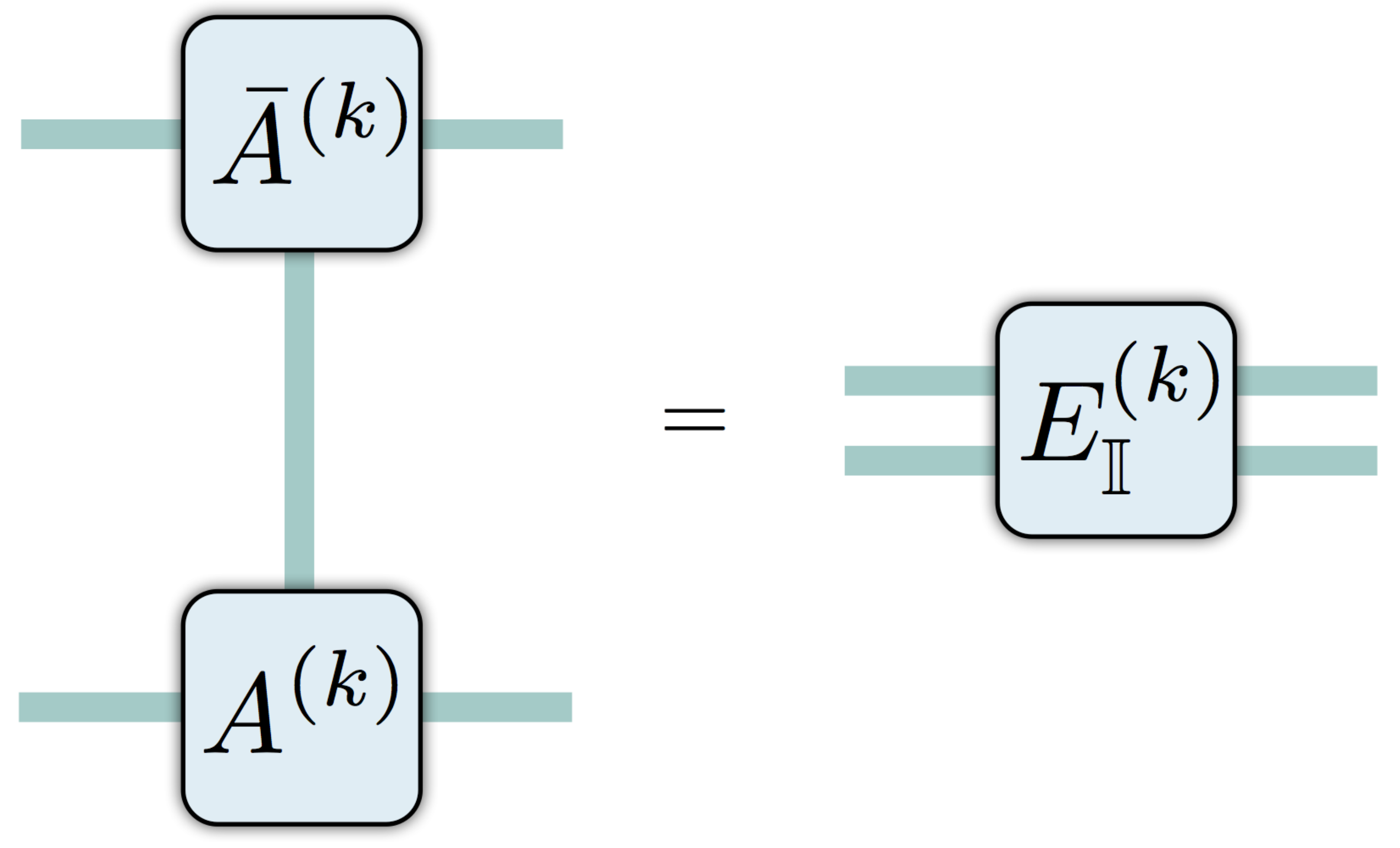}.
\end{figure}

We can progress until we come to the sites on which the local observable $O$ is supported. But of course, we can still contract all of the physical indices and treat it as
one new tensor, to get $E_O$,
\begin{figure}[h!]
 \centering
\includegraphics[width=0.35\textwidth]{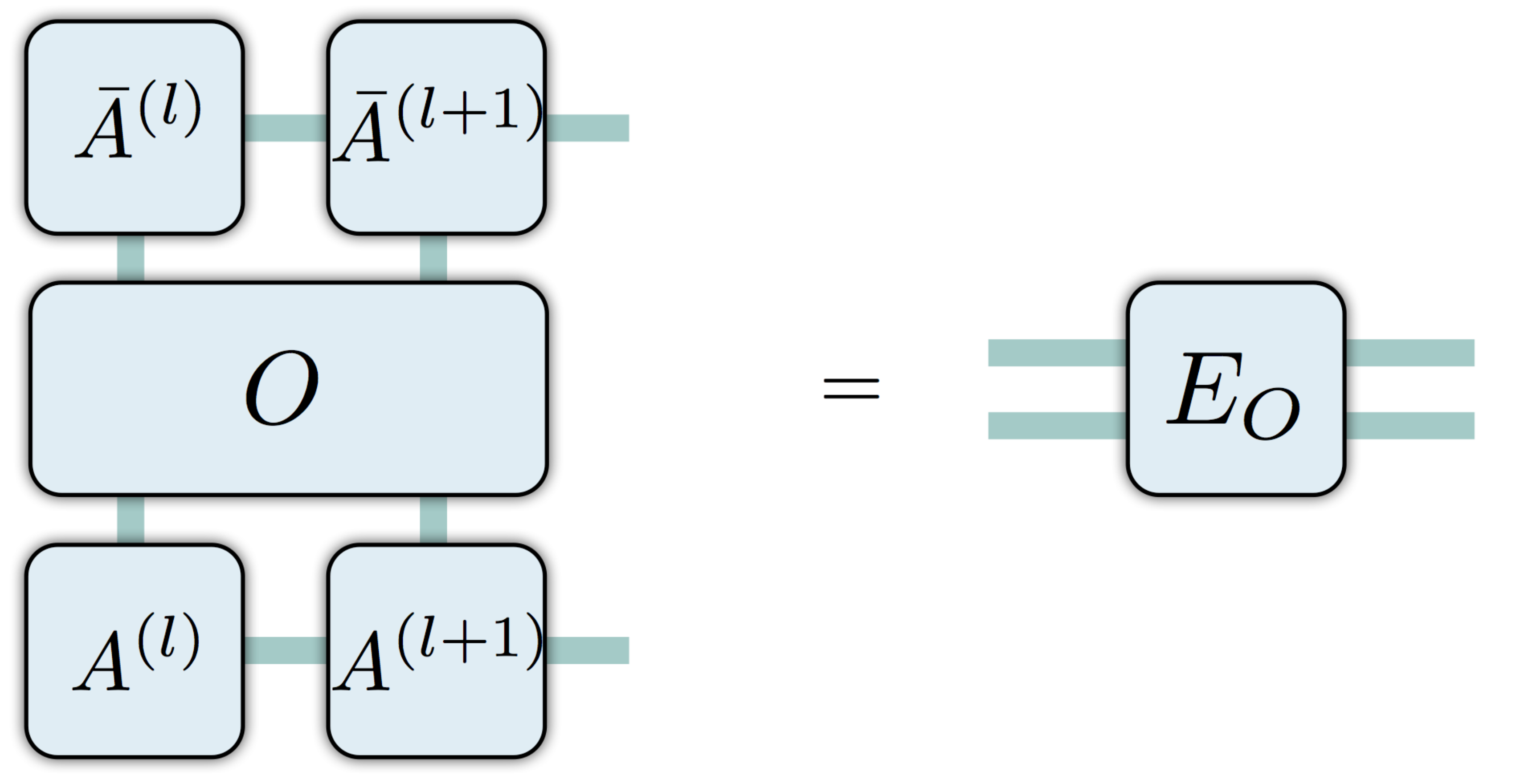}.
\end{figure}

Finally, the {\it right boundary condition} can be captured as
\begin{equation}
	R_{\alpha,\beta}=  \sum_{j=1}^d A_{\alpha;j}^{(n)} \bar{A}_{\beta;j}^{(n)},
\end{equation}
graphically
\begin{figure}[h!]
 \centering
\includegraphics[width=0.248\textwidth]{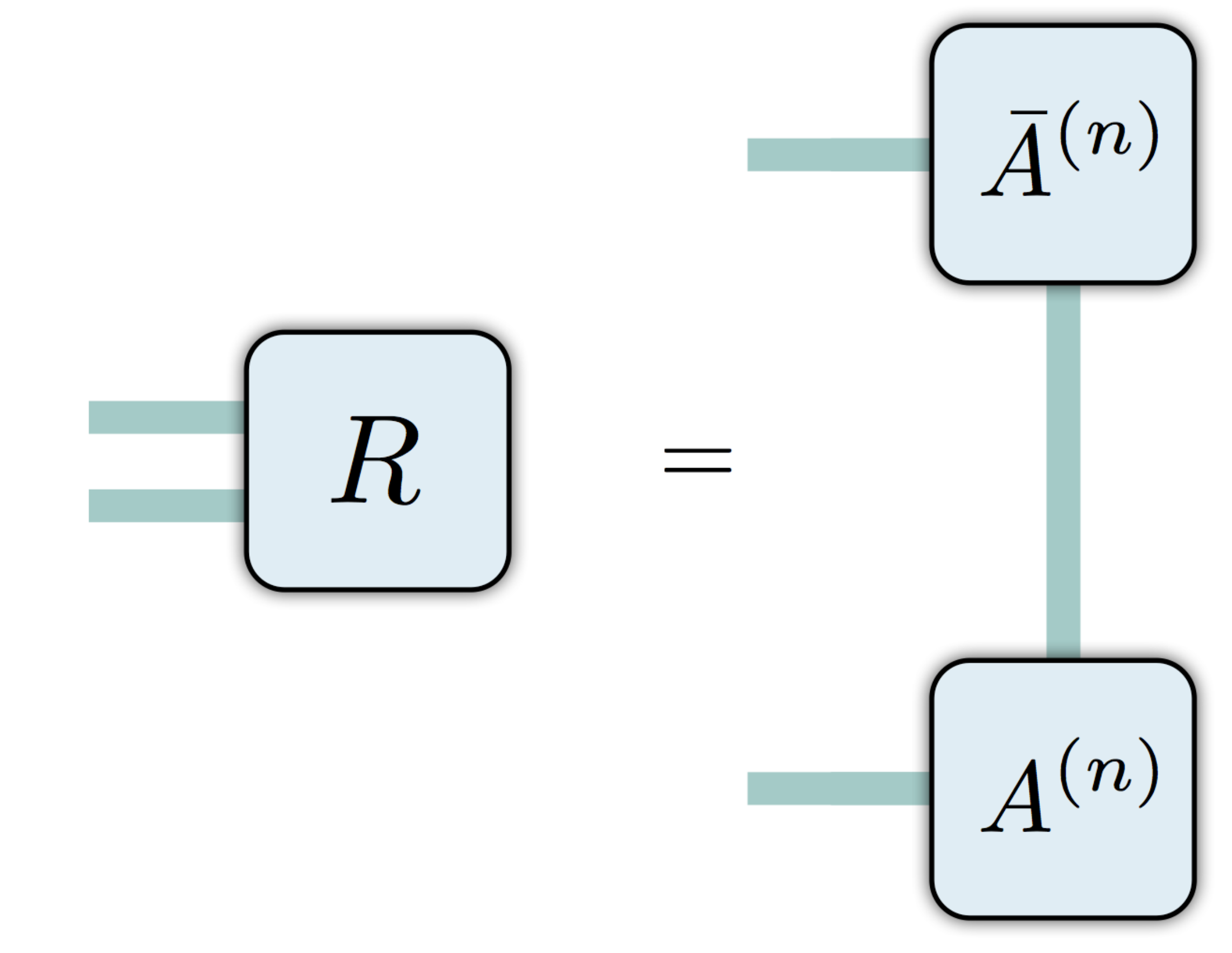} .
\end{figure}

So the above graphical representation can also be sequentially read as a representation of the expression
\begin{equation}
	\langle\psi| O|\psi\rangle = L E_\id^{(2)} E_\id^{(3)}\dots E^{(l-1)}  E_O E_\id^{(l+2)}\dots E_\id^{(n-1)} R.
\end{equation}

Of course, not only all these operators can be efficiently computed, but also the product be performed with the same effort as it is needed to 
multiply a vector from $\cc^{D^2}$ with matrices from $\cc^{D^2\times D^2}$, namely $O(d D^4)$. Since there are $n$ sites, the total effort to compute a single
expectation value can be bounded by
$O(n d D^4)$. For a local Hamiltonian with $O(n)$ terms $H$ this amounts to an effort $O(n^2D^4)$ to compute $\langle \psi| H |\psi\rangle$. That is to say, one can 
efficiently compute expectation values of local Hamiltonians.
It is easy to see that this can be further improved to $O(n D^3)$, by using an appropriate gauge, to be specified below, and altering the contraction order still.

\subsubsection{Decay of correlations in infinite matrix product states}

We have seen above that MPS capture ground states of local one-dimensional gapped models well. As such, one should expect that also common features of 
such models are appropriately reflected. In particular, one should expect correlation functions to decay exponentially with the distance in the lattice. In fact, the object we need
in order to see this is the transfer operator encountered earlier. We stick to the situation of a infinite translationally invariant MPS, so the transfer operator 
\begin{equation}
	E_\id	 = \sum_{j=1}^d (A_j \otimes \bar A_j),
\end{equation}
graphically
\begin{figure}[h!]
 \centering
\includegraphics[width=0.31\textwidth]{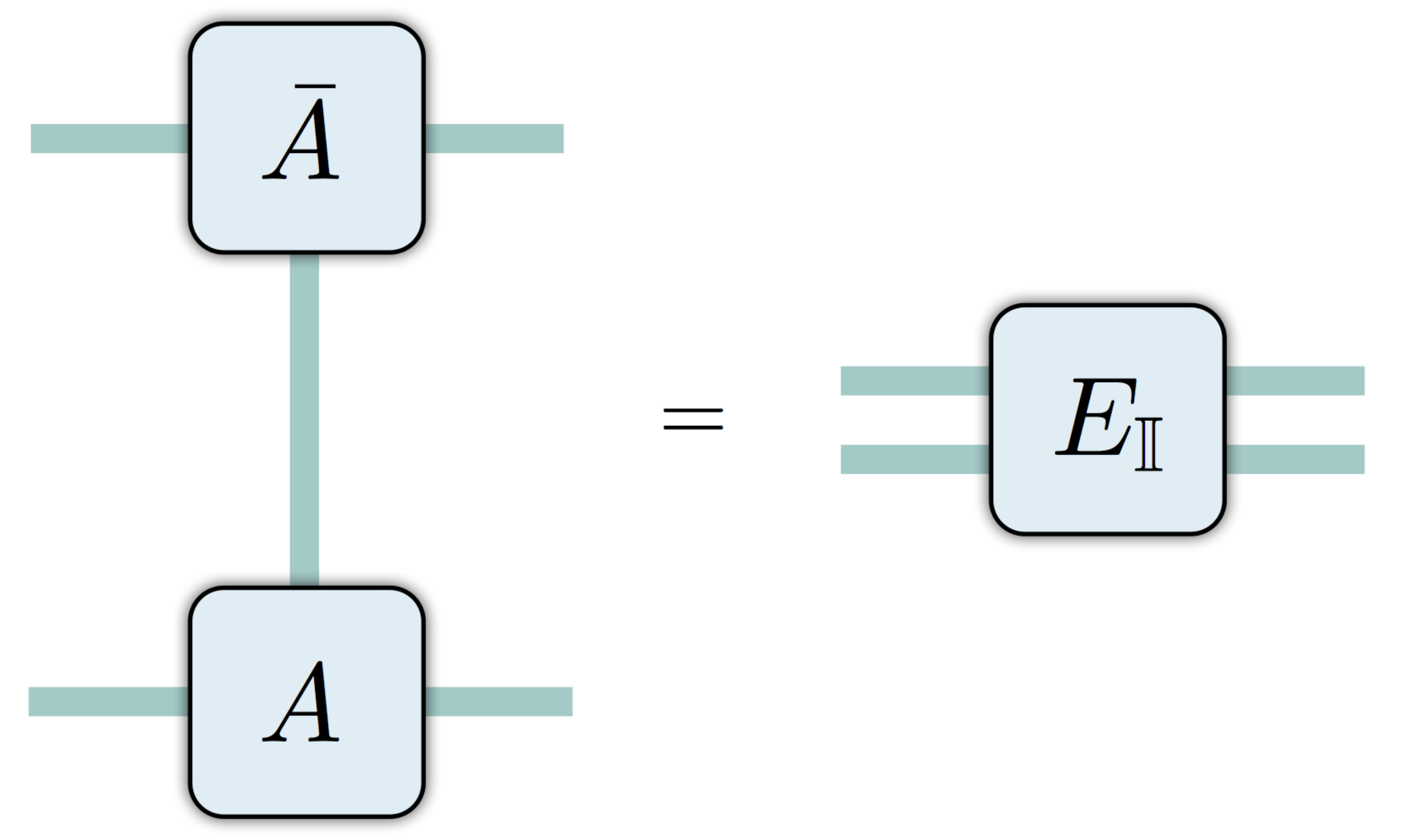},
\end{figure}

will not
carry a site index.We will consider correlation functions between two sites that we label $A$ and $B$ as above. The observables supported on $A$ and $B$ are again
$O_A$ and $O_B$. In line with the above definitions, we set
\begin{equation}
	E_{O_A} = \sum_{j,k=1}^d \langle k | O_A | j\rangle (A_j\otimes \bar{A}_k),
\end{equation}
graphically,
\begin{figure}[h!]
 \centering
\includegraphics[width=0.315\textwidth]{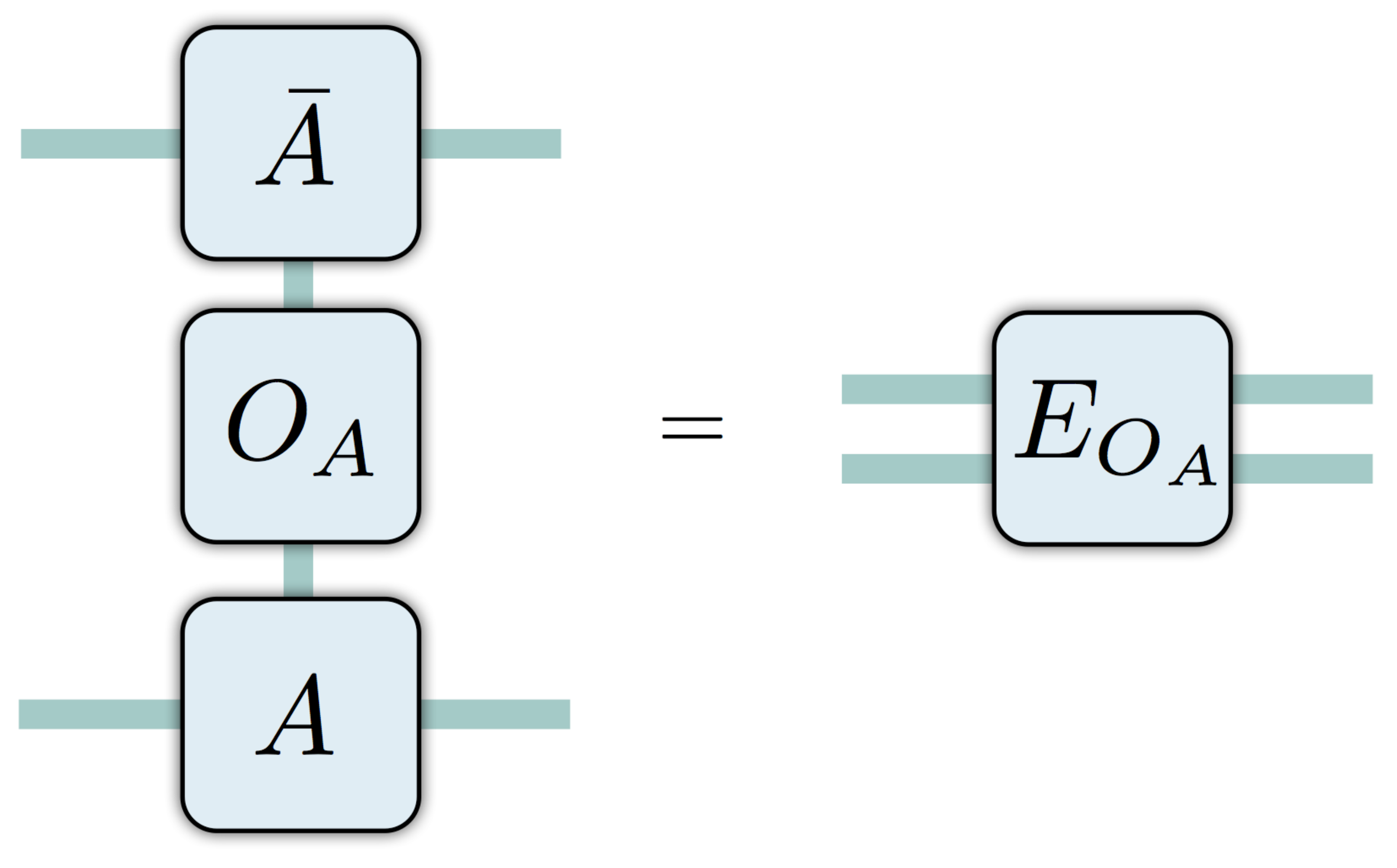},
\end{figure}

and similarly for $O_B$. Then the correlation function is obtained as
\begin{equation}
	\langle O_A O_B\rangle = \frac{\tr (E_{O_A} E_\id^{\dist(A,B)-1} E_\id^{n-\dist(A,B)-1})}{\tr( E_\id^n)}.
\end{equation}
We are now specifically interested in the limit $n\rightarrow\infty$ in {\it infinite translationally invariant MPS}.
We have that
\begin{equation}
	E_\id^k =  |r_1\rangle\langle l_1 |+ \sum_{j=2}^{D^2} \lambda_j^k |r_j\rangle\langle l_j |, 
\end{equation}
where $\lambda_1=1,\lambda_2,\dots, \lambda_{D^2}$ are the eigenvalues of $E_\id$ in non-increasing order and $|r_j\rangle$ and $\langle l_j|$ the 
respective right and left eigenvectors. To assume that the largest eigenvalue $\lambda_1= \|E_\id\|=1$ does not restrict generality -- this merely amounts to a
rescaling. We also assume that this is the unique eigenvalue that takes the value $1$. Then, in the limit $n\rightarrow \infty$, 
\begin{equation}
	\langle O_A O_B\rangle = \langle l_1 | E_{O_A} E_{\id}^{\dist (A,B)-1} E_{O_B}|r_1\rangle.
\end{equation}
This is nothing but
\begin{equation}
	\langle O_A O_B\rangle = \langle l_1 | E_{O_A}|r_1\rangle\langle l_1 | E_{O_B} | l_1\rangle + \sum_{j=2}^{D^2} \lambda_j^{\dist (A,B)-1} \langle l_1 | E_{O_A}|r_j\rangle
	\langle l_j| E_{O_B}| l_1\rangle ,
%
\end{equation}
where the first term can be identified as $\langle O_A\rangle\langle O_B\rangle$.
This means that $|\langle O_A O_B\rangle -\langle O_A\rangle\langle O_B\rangle|$ decays exponentially in the distance $\dist (A,B)$, and the correlation length
$\xi>0$ is given by the ratio of the second largest $\lambda_1$ to the largest $\lambda_1=1$ (here taken to be unity) eigenvalue of $E_\id$, so
\begin{equation}
	\xi^{-1} = - \log | \lambda_2|.
\end{equation}
This is a very interesting observation: The decay of correlations is merely governed by the spectral gap between the two largest eigenvalues of the transfer operator.
All other details of the transfer operator do not matter asymptotically as far as the decay of correlations is concerned. This also means that whenever this gap is not vanishing,
correlation functions always decay exponentially. Positively put, this may be taken as yet another indication that MPS represent ground states of gapped models well 
(for which correlation functions are similarly decaying). 
Higher order correlation functions of the form $\langle O_A O_B O_C\rangle$ and so on can also be efficiently computed from the MPS representation in the same way.
There is an interesting structural insight related to this: 
In order to fully specify an MPS of any bond dimension, generically, 
the collection of all correlation functions of order up to merely three need to be specified \cite{Wick}, and not of all orders, as one would naturally expect.

In fact, MPS cannot represent algebraically decaying correlation functions, even though one should not forget that
for sufficiently large bond dimension, MPS can well approximate states with algebraically decaying correlation functions as well.
One might be tempted to think that this is different
whenever the gap is vanishing, meaning that whenever $\lambda_2=1$ and $\lambda_1=1$ are degenerate. This is not so, however. Then one rather encounters 
constant contributions to correlation functions (we will learn about an example of this form in form of the GHZ state
below).

\subsubsection{Computation of scalar products between matrix product states}

It is an interesting exercise to verify that the scalar product of two different (non-translationally invariant, open boundary condition) 
MPS of bond dimension $D$ can be computed with an effort $O(n d D^3)$. This fact can be quite helpful in practical computations.

\subsubsection{Density-matrix renormalisation method in several invocations}

\index{Density matrix renormalisation group method}
The workhorse of the numerical study of strongly correlated one-dimensional systems is the {\it DMRG method}, introduced in the seminal Ref.\
\cite{White}. Albeit this was not the way this method has originally been formulated, it has become clear \cite{Rommer,Kluemper} that it can be viewed
as a variational method over MPS: In one way or the other, one varies over MPS state vectors $|\psi\rangle $ of a given bond dimension, until 
for a given local Hamiltonian $H=\sum_{j\in V }h_j$ a good approximation of
\begin{equation}\label{mini}
	\min \frac{\langle \psi| H |\psi \rangle }{\langle\psi|\psi\rangle}
\end{equation}
is reached. We can describe the MPS by polynomially many (in fact in $n$ linearly many) parameters that can 
be stored in a computer, and we can efficiently compute expectation values. Since the optimisation over all parameters at once amounts to a non-convex global
optimisation problem (and is hence infeasible), this task is broken down to a sequential updating of the matrices $\{A^{(k)}_{j_k}\}$ of the MPS. For example,
starting from randomly picked matrices in the MPS, if one holds all matrices except those $\{A^{(j)}_1,\dots, A^{(j)}_d\}$ of a site $j$ fixed, then one can write the optimisation problem of Eq.\ (\ref{mini})
as a minimisation over
\begin{equation}
	E :=\frac{\langle \psi| H |\psi \rangle }{\langle\psi|\psi\rangle} = \frac{\langle A^{(j)} |K_1| A^{(j)} \rangle }{\langle A^{(j)} |K_2| A^{(j)} \rangle}
\end{equation}
with $| A^{(j)} \rangle$ denoting the vectorized forms of the matrices and $K_1$ and $K_2$ being the kernels of the respective quadratic forms. This is not only
a convex quadratic
optimisation problem, but in fact an eigenvalue problem $K_1|  A^{(j)} \rangle = E K_2  |A^{(j)} \rangle$. In this
way, by `sweeping through the lattice', going from one site to the next and coming back, until convergence is reached,
one can find ground state properties practically essentially up to machine precision. 
In practical, often surprisingly few sweeps are needed to reach a stable minimum, even if one starts off with random MPS.
See Refs.\ \cite{Scholl,SchollwoeckAge,VerstraeteBig} for reviews\footnote{As a side remark, strictly speaking, it is not guaranteed by this
procedure that one really obtains the global minimum when performing local variations. In fact, practically one may get stuck, and it can be 
beneficial to insert manually artificial fluctuations \cite{White}. In practice, in the context discussed in these lecture notes,
this is usually not much of a problem, however.
The {\it computational complexity of actually finding the optimal MPS}, given a fixed family of Hamiltonians and a given bond dimension,
has been addressed in Refs.\ \cite{SchuchComplexity,SchuchNP,NP}.}.

Having said that, there are, needless to say, many different methods of how to proceed and many different ways in which 
one can improve this basic method. To start with,
a clever use of the gauge (as we will discuss later) is crucial to arrive at a practical implementation avoiding {\it ill-conditioned matrices} along the way. 
Then, one does not have to vary over one set of matrices per step,
but can vary over pairs of matrices, in a {\it double variational site}, leading in particular to a faster convergence and a better error control.
One can employ instances of {\it time-evolving block decimation (TEBD)} \cite{VidalTEBD,Daley,WhiteFeiguin}
in imaginary-time evolution (see the next subsection), or the {\it time-dependent variational
principle} \cite{TimeDependentVariationalPrinciple} avoiding Trotter errors. Some variants of DMRG avoid errors from finite-size scaling by directly referring to 
infinite MPS (possibly with a broken symmetry with a finite period), such as the {\it iDMRG method} \cite{iDMRG1,iDMRG2} or {\it iTEBD} \cite{iTEBD}.
A refined concept of time evolution is also constituted by {\it concatenated tensor networks} \cite{ConcatenatedTensorNetworks}.

\subsubsection{Matrix-product operators and mixed states}

Tensor network states can not only represent pure states, but mixed quantum states as well. A {\it matrix product operator} $O\in{\cal B}((\cc^d)^{\otimes n})$
relates to a tensor network\begin{equation}
	O=\sum_{j_1,\dots, j_n=1}^d \sum_{k_1,\dots, k_n=1}^d  \tr(A^{(1)}_{j_1,k_1}\dots A^{(n)}_{j_n,k_n} )|j_1,\dots, j_n\rangle\langle k_1,\dots, k_n|,
\end{equation}
\index{Matrix product operators}
These operators contain mixed quantum states (and also other operators which are not positive in the sense that eigenvalues can be negative; 
in fact, checking positivity of such an matrix product operator is not
straightforward).  Graphically, they can be represented as tensor networks
\begin{figure}[h!]
 \centering
\includegraphics[width=0.47\textwidth]{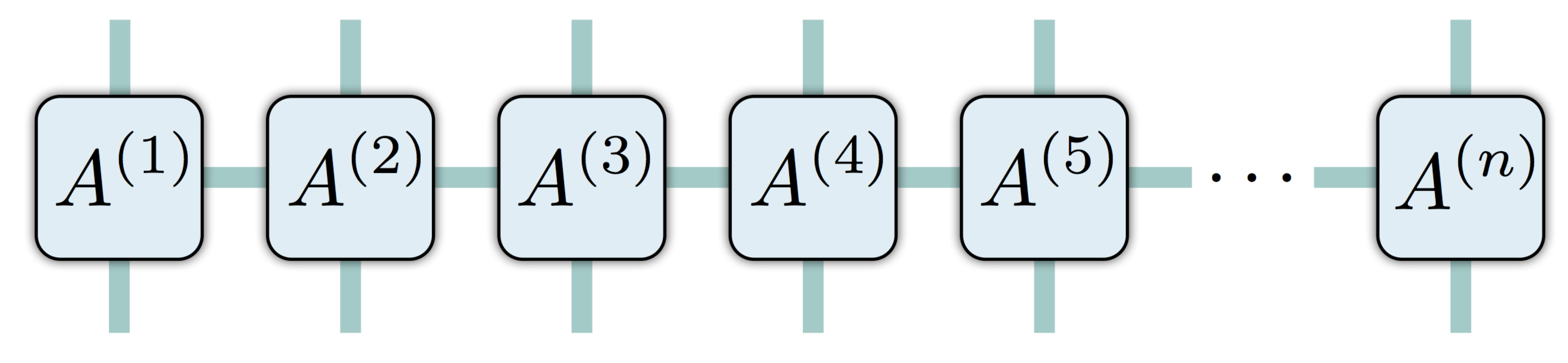}.
\end{figure}

One can now devise algorithms that represent mixed states such as Gibbs states 
\begin{equation}
\rho_\beta:=	e^{-\beta H}/\tr(e^{-\beta H})
\end{equation}
in variants of the DMRG algorithm \cite{Mixed,Zwolak,ZnidaricPP08}.
There is a second picture capturing mixed states, which are obtained from MPS with a special structure by taking a partial trace over the purifying degrees of freedom, here
depicted in the form of an isomorphism between purified MPS and positive instances of MPO \cite{BarthelMPO}, graphically

\begin{figure}[h!]
 \centering
\includegraphics[width=0.47\textwidth]{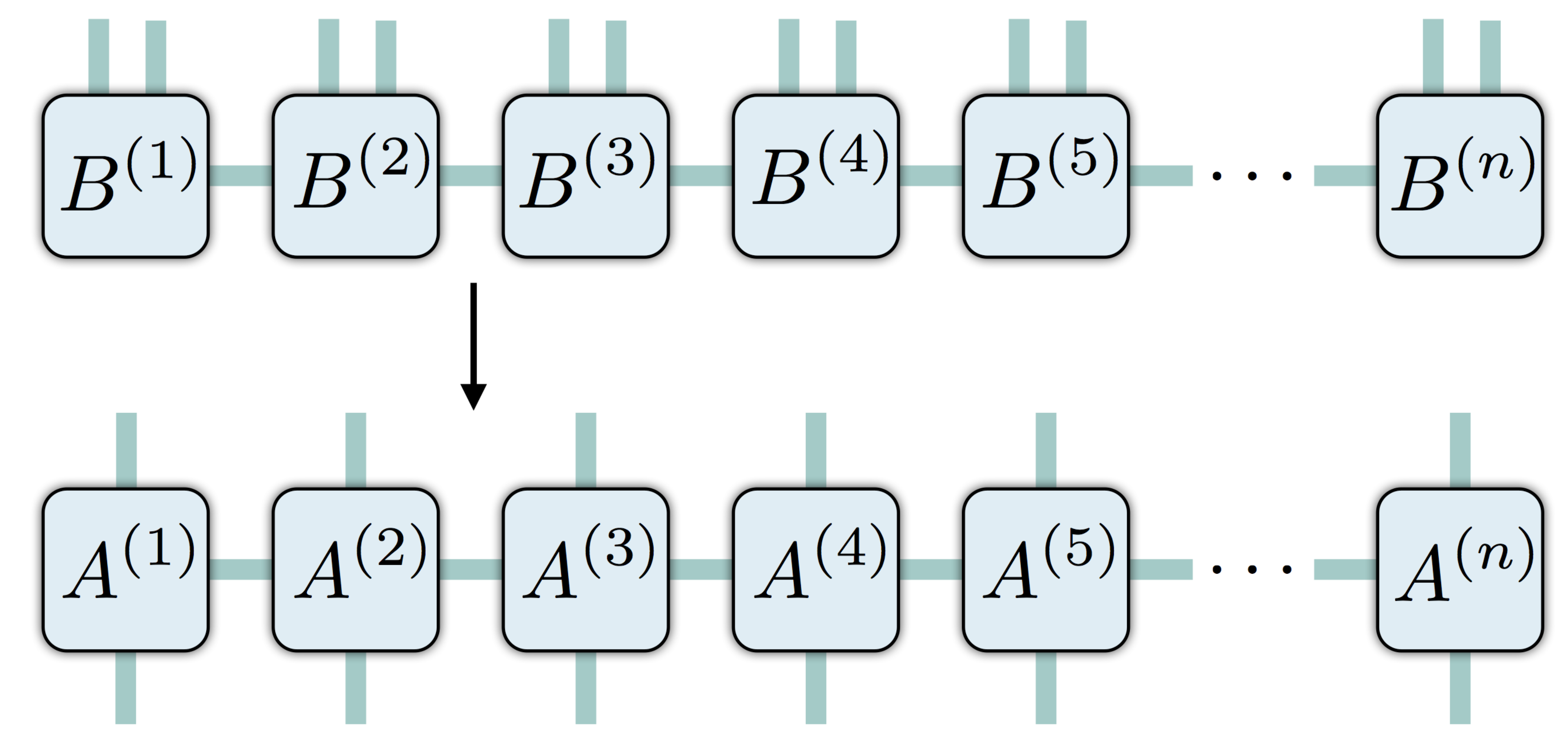} .
\end{figure}

\subsubsection{Time evolution}

Time evolution algorithms provide numerical computations of expectation values 
\begin{equation}
	\langle O_A\rangle(t) :=\langle e^{\i tH } O_A e^{- \i tH }\rangle 
\end{equation}
for a local Hamiltonian $H$ and a local observable $O_A$ supported on several sites. Basically, there are two major strategies known to proceed here,
as far as MPS simulations are concerned:
\begin{itemize}
\item On the one hand, one can decompose the Hamiltonian $H=H_1+H_2$ into a sum of an even and an odd part, such that both Hamiltonians contain non-overlapping Hamiltonian terms only. One can then approximate $e^{-\i H t}$ by $(e^{\i t H_1 t/k} e^{\i tH_2 t/k})^m$ for a suitable large $m\in \nn$, with good control of errors. 
The {\it time-evolving block decimation (TEBD)} \cite{VidalTEBD,Daley,WhiteFeiguin} and variants thereof make use of that idea.\index{Time-evolving block decimation method}
\item On the other hand, one can make use of the {\it time-dependent variational principle} \cite{TimeDependentVariationalPrinciple}, which relies
on the variational manifold of uniform MPS mentioned above and which avoids Trotter errors altogether.
\end{itemize}
Such methods have provided significant new insights into {\it non-equilibrium dynamics of strongly correlated quantum systems} and have given new impetus to the study
old questions of equilibration, thermalisation. Specifically, quenched quantum lattice models have been extensively studied,
resulting in much progress on questions of far-from-equilibrium dynamics (see, e.g., Refs.\ \cite{Kollath,QuenchDMRG,Bloch,Impurity}).
\index{Time-dependent variational principle}
\index{Quenched dynamics}

For short times, starting from clustering initial conditions,
these methods again provide very reliable information about the evolution of the strongly correlated many-body systems. The reason is again rooted in the above
mentioned Lieb-Robinson bounds: One can show that under local Hamiltonian dynamics, for any fixed time, an area law for Renyi entanglement
entropies holds true
\cite{Quench}. Hence, an efficient approximation of the true time evolved states with an MPS is possible. The prefactor in the area law grows with time, however,
leading to the situation that the entanglement effectively increases linearly in time in worst case \cite{SchuchQuench}. That is, for long times, one cannot
capture time evolution of quantum many-body systems with MPS: One hits the `barrier of entanglement growth'. So-called {\it folding method} that contract the 
emerging tensors in a different way soften this problem to some extent \cite{Folding}. Still, to grasp long time dynamics is infeasible, and it is one of the interesting
open problems to see
to what extent this challenge can be overcome.

\subsection{Parent Hamiltonians, gauge freedom, geometry, and symmetries}


At this point, it should have become clear that MPS are primarily meant to approximate natural states, specifically ground states of gapped one-dimensional
local Hamiltonian models. Yet, a natural question that arises at this point is whether there are meaningful Hamiltonians that have an MPS as their
exact 
ground state.
We will now take a look at this question. Starting from this observation, we will hint at the insight that indeed, MPS (and more generally tensor network states)
are by no means only powerful tools in numerical simulations, but also give rise to a very versatile tool in analytical considerations. Since so many questions can be
fully analytically assessed (in contrast to many actual Hamiltonian problems), an entire research field has emerged of `studying condensed matter physics in MPS world'.
For example, complete classifications of quantum phases have been given  \cite{ClassificationSchuch,ClassificationWen,ClassificationPollmann}, 
new instances of {\it Lieb-Schultz-Mattis theorems} proven \cite{MPSLSM}, {\it fractional magnetisation} considered \cite{Fractional}
{\it phase transitions of arbitrary order} identified \cite{MPSOrderPhaseTransitions}, or a {\it `Wick's theorem'} for MPS be formulated \cite{Wick}.
It will be beyond the scope of this book chapter to give a comprehensive overview of this development.

\subsubsection{The AKLT and the Majumdar Gosh models}
\index{AKLT model}

Surely, any product state is an MPS with bond dimension $D=1$, so every state that has a product state as an exact ground state will provide an example of that sort.
This is not very exciting yet, however. 
\begin{itemize}
\item The most famous model that has an MPS ground state with a bond dimension different from $D=1$
is the {\it AKLT model}, 
named after Affleck, Kennedy, Lieb, and Tasaki \cite{AKLT}. 
Here, the bond dimension is $D=2$ while
the local physical dimension is $d=3$: It is a spin-$1$ chain. There are many ways to progress to see that this model indeed has an MPS as its ground state.
One particularly transparent way makes use of the above projected entangled pair picture -- historically, this is actually where this construction originates from.
The linear projections are all the same for all sites and taken to be
\begin{equation}
	P=\Pi_{S=1}(\id\otimes \i Y)
\end{equation}
where $\Pi_{S=1}$ is the projection onto the spin-$1$ subspace of two sites, and $Y$ is the Pauli matrix. This might look a bit awkward at first, but it is a sanity check that 
it takes a state defined on two spin-$1/2$ systems (with Hilbert space $\cc^2\otimes \cc^2$) to one spin-$1$ state (defined on $\cc^3$). The Pauli matrix 
is just added here for the convention that we always start in the maximally entangled state as defined in Eq.\ (\ref{maxent}). So what this linear map essentially does is that it takes bonds prepared in singlets
and projects them into the $S=1$ subspace. This surely gives rise to a valid MPS with state vector $|\psi\rangle$ .

Why does this help to construct a Hamiltonian that has this MPS as the ground state? We can simply take as the local Hamiltonian term $h_j= \Pi_{S=2}$, so that surely
$h_j |\psi\rangle=0$ for all $j$. That is to say, the Hamiltonian terms are the projections onto the $S=2$ subspace. For the same reason,
\begin{equation}\label{aklt}
	H|\psi\rangle = \sum_{j} h_j |\psi\rangle =0
\end{equation}		
(mildly disregarding some fine-print on the boundary conditions).
But we have that $h_j\geq 0$, so that each term has non-negative eigenvalues, which means that $|\psi\rangle$ must be a ground state vector. There is some fine print involved here,
as strictly speaking, we have only seen that it constitutes a valid ground 
state vector, not really whether it is a unique one. This can be shown to be true, however, by identifying the 
Hamiltonian in Eq.\ (\ref{aklt}) as the so-called parent Hamiltonian of the given MPS, a concept we will turn to later.

How does the resulting Hamiltonian then look like? A moment of thought reveals that it can be written as
\begin{equation}
	h_j = \frac{1}{2}S^{(j)} \cdot S^{(j+1)} + \frac{1}{6} (S^{(j)}\cdot S^{(j+1)})^2+\frac{1}{3},
\end{equation}
The matrices of the MPS are found to be $A_1= X$, $A_2= (X+\i Y)/\sqrt{2} $, $A_3= -(X-\i Y)/\sqrt{2} $.
In fact, one of the motivations to study the AKLT model is also the close resemblance to the {\it spin-1 Heisenberg model} the Hamiltonian of which 
has local terms
\begin{equation}\label{Heisenberg}
	h_j = J S^{(j)} \cdot S^{(j+1)} 
\end{equation}
for some $J\in \rr$. This model is important for numerous reasons. It is also connected to the famous {\it Haldane conjecture}, which 
states that integer-spin anti-ferromagnetic Heisenberg chains are gapped \cite{Haldane}.

\item Another important local Hamiltonian that has an exact MPS ground state is one of the {\it Majumdar-Gosh model} \cite{Majumdar}, 
a nearest-neighbour 
spin-$1/2$ chain of local dimension $d=2$ with Hamiltonian
\begin{equation}
	H=\sum_{j}\left(
	2 \sigma^{(j)} \cdot \sigma^{(j+1)} + \sigma^{(j)} \cdot \sigma^{(j+2)}
	\right),
\end{equation}
where $\sigma$ is the vector of Pauli matrices. It turns out that its ground state can be represented with matrices of bond dimension $D=3$.
\index{Heisenberg model}
\end{itemize}


\subsubsection{Gauge freedom and canonical forms}\label{canonical}

An MPS is uniquely defined by the matrices defining it, but the converse is not true: There is more than one set of matrices that give rise to the same pure state.
Since
\begin{equation}
	A_{j_k}^{(k)} A^{(k+1)}_{j_{k+1}} = A_{j_k}^{(k)} X X^{-1} A^{(k+1)}_{j_{k+1}} 
\end{equation}
for any $X\in Gl(D,\cc)$, graphically represented as
\begin{figure}[h!]
 \centering
\includegraphics[width=0.5\textwidth]{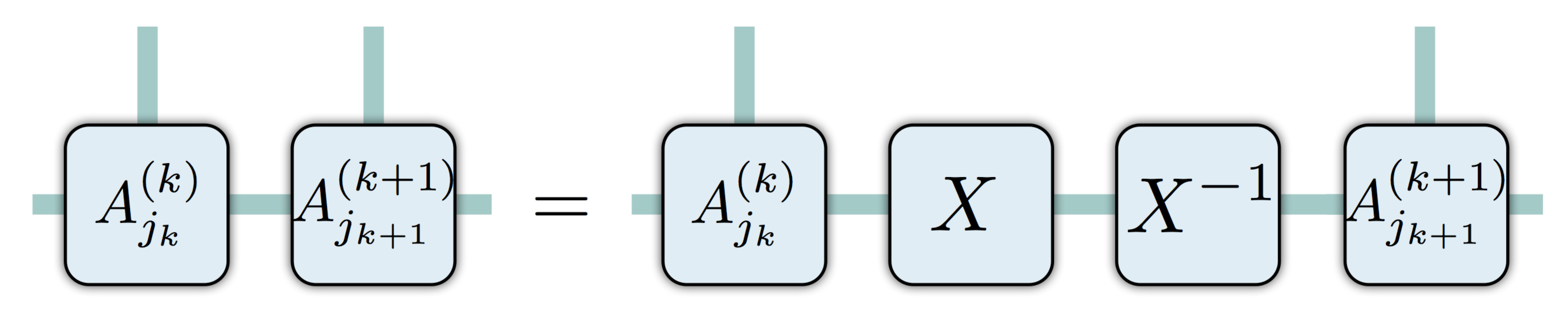},
\end{figure}

one can introduce between any pair of MPS matrices an invertible matrix. Choosing a specific $X$, or rather choosing such matrices for an 
entire MPS, amounts to choosing a so-called 
{\it gauge}. This insight, basic as it is, is very helpful in many circumstances. {\it Canonical forms}, so special forms of MPS that are
particularly simple and which can be achieved by picking a suitable gauge, are very useful in analytical considerations. They are also helpful in numerical methods: For example,
the above reduction of the effort when computing expectation values from the naive $O(n^2 D^4)$ to $O(n D^3)$ is also partially due to picking the right gauge.
For MPS with open boundary conditions and bond dimension $D$, one can, for example, always pick a gauge in which
\begin{eqnarray}
	\sum_{j} A_j^{(k)} (A_j^{(k)})^\dagger &=& \id,\\
	\sum_{j} (A_j^{(k)})^\dagger \Lambda^{(k-1) } A_j^{(k)}  &=& \Lambda^{(k)},\\
	\Lambda^{(0)}=\Lambda^{(n)}=1
\end{eqnarray}
and each $\Lambda^{(k)}\in \cc^{D\times D}$ 
for $k=1,\dots, n-1$ is diagonal, positive, has full rank and unit trace. This can be shown by a successive singular value decomposition. For periodic
boundary conditions, finding appropriate normal forms is more involved, see Ref.\ \cite{MPSSurvey}.

\subsubsection{Injectivity and parent Hamiltonians}

We have seen that the AKLT model has a unique MPS as its ground state, and so does the Majumdar Gosh model. Indeed, these are examples of a more
general framework that goes back to the seminal work of Ref.\ \cite{FCS}, where the question has been solved for infinite systems. Here, we ask, for a finite MPS, 
when is it the unique ground state of a gapped local Hamiltonian? The latter is usually called {\it parent Hamiltonian} of the MPS. 
Let us assume we have a MPS state vector in the canonical form of Subsection \ref{canonical}
\begin{equation}
	|\psi\rangle = \sum_{j_1,\dots, j_n=1}^d A^{(1)}_{j_1}\dots A_{j_n}^{(n)}|j_1,\dots, j_n\rangle.
\end{equation}
We assume now that we can group the constituents such that the grouped MPS, now no longer of $n$ but of $m$ new sites of larger local dimension, to get
\begin{equation}
	|\psi\rangle = \sum_{j_1,\dots, j_m} B^{(1)}_{j_1}\dots B_{j_m}^{(m)}|j_1,\dots, j_m\rangle.
\end{equation}
We require that each sets of matrices $\{B^{(k)}_{j_k}\}$ has the property that it generates the respective entire space of matrices.\footnote{Let $\{A^{(k)}_{j_k},\dots, A^{(k+L-1)}_{j_{k+L-1}}\}$ correspond to the block called $B^{(s)}$  for some suitable $s$, then the above means that
the map
$\Gamma_L: Gl(D,\cc)\rightarrow (\cc^d)^{\otimes L}$
with 
$\Gamma_L: X\mapsto \sum_{j_k,\dots, j_{k+L-1}=1}^d 
	\tr(
	X A_{j_k}\dots A_{j_{k+L-1}}
	) |j_1,\dots, j_L\rangle$is {\it injective} \cite{MPSSurvey}. If for a state vector such an $L$ can be found, then the state vector is called injective.
}
%
Then one can find local Hamiltonian terms $\{h_j\}$ each supported on $L+1$ sites of the original lattice such that 
 $|\psi\rangle $ is the unique ground state vector of 
$H= \sum_j h_j$.
This is in fact a generalisation of the above idea that we encountered in the AKLT model (strictly speaking, there injectivity sets in at $k=2$, so one would in principle
arrive at a $3$-local Hamiltonian, but one can show that this Hamiltonian and the given nearest-neighbour Hamiltonian of the AKLT model are in fact identical). The
existence of a gap can for finite systems be shown in essentially the same way as for infinite finitely correlated states \cite{FCS}.
The idea that one can (under the above mild technical conditions) 
find a gapped local Hamiltonian of which the MPS is the unique ground state is a very powerful one in analytical uses of MPS.
\index{Parent Hamiltonian}

\subsubsection{Group symmetries}

Significant progress has been made in recent years in the study tensor networks -- and specifically MPS -- under symmetry constraints. Here the emphasis is
pushed from the study of symmetries in Hamiltonians to those of states, but there is a close connection (making use of the concept of a parent Hamiltonian of the previous
subsection): If an MPS state is invariant under a representation of a group, then one can choose its parent Hamiltonian to be invariant under the same representation.
 
One of the key features of a translationally invariant state vector $|\psi\rangle$ on $n$ sites
is the symmetry group $G$ under which it is invariant: This is the group for which
\begin{equation}
	u_g^{\otimes n} |\psi\rangle  = e^{\i \phi_g}|\psi\rangle,
\end{equation}
where $g\in G$ and $u_g$ is a unitary representation on ${\cal H}$. It turns out that for translationally invariant MPS that fulfill the 
injectivity condition this symmetry is reflected also by a group symmetry in the tensors that define the MPS:
The action of the unitary $u_g$ on the physical index corresponds to an action of a $V_g$ on the virtual level. More specifically,
$u_g P = P(V_g \otimes  \bar {V}_g )$ for the linear operators $P$ defining the MPS ($P=P^{(j)}$ for all $j$ in a translationally invariant ansatz).
This picture has shed new light on the concept of
{\it string order} \cite{StringOrder}. 
It also plays an important role in the {\it classification of phases} \cite{ClassificationSchuch,ClassificationWen,ClassificationPollmann}, when
 two gapped systems are defined to be in the {\it same phase} if and only if they can be connected by a smooth path of gapped local Hamiltonians.

\subsubsection{Manifold of matrix product states}

There is an interesting geometrical structure associated with (translationally invariant) MPS. We have seen that there is a gauge freedom in MPS, leading to an 
over-parametrisation. Due to this redundancy in parametrisation, MPS have the structure of a principal fiber bundle. The bundle space corresponds to the entire
parameter space, that is, the collection of all tensors associated with the physical sites. The base manifold, in turn, is embedded in in the Hilbert space.
This geometrical structure is fleshed out in detail in Ref.\ \cite{JuthoManifold}.


\subsection{Applications in quantum information theory and quantum state tomography}
\subsubsection{Matrix product states in metrology}

Many multi-particle entangled states that are interesting in the context of quantum information theory and metrology can also be represented as matrix product
states. The well-known {\it Greenberger-Horne-Zeilinger (GHZ) state} with state vector
\begin{equation}
	|\psi\rangle= (|0,\dots, 0\rangle + |1,\dots, 1\rangle)/\sqrt{2},
\end{equation}
for example, can be written as a MPS with bond dimension $D=2$ and $A_1=|0\rangle\langle 0|$ and $A_2= |1\rangle\langle 1|$.
For practical applications in metrology, GHZ states are too fragile with respect to noise, and other states which can also be 
represented as MPS are more useful \cite{MPSMetrology}.

\subsubsection{Matrix product states in measurement based quantum computing}

{\it Cluster states} are also MPS: These states are an important class of states in quantum information theory, most prominently featuring (in their two-dimensional instance)
in ideas of {\it measurement-based quantum computing} \cite{Oneway,Models1}: 
This is the idea of performing quantum computing without the need of actual unitary control over arbitrary pairs
of constituents, but rather by sequentially (and adaptively) measuring single sites. Since read-out has to be done anyway at some stage even in the circuit model, 
this amounts to an appealing picture of quantum computing. All the entanglement `consumed' in the course of the computation is already present in the initial, rather simple,
resource state. Cluster states are an instance of the more general {\it graph states} \cite{Graphs}, which constitute a helpful theoretical `laboratory': They can be
viewed as prepared in the following way: One starts with preparing the vertices in a lattice $V$ in $|+\rangle = (|0\rangle+|1\rangle)/\sqrt{2}$ and applying {\it controlled phase gates}
\begin{equation}
	|j,k\rangle\mapsto |j,k\rangle e^{\i \delta_{j,1}\delta_{k,1}\phi}
\end{equation}
to neighbouring systems for $\phi\in [0,2\pi)$; 
in the original definition, the 
phase $\phi=\pi$ is chosen.\footnote{In fact, even states that can be prepared by applying arbitrary non-local
phase gates associated to any interaction graph applied to an arbitrary MPS can be efficiently contracted. This is possibly by suitably defining 
transfer operators that absorb the phases in such a way that the long-range entanglement is not an obstacle to an efficient contraction. 
The schemes arising from this variational set of states are referred to as
{\it renormalisation schemes with graph enhancement} \cite{Rage}. Such states are efficiently contractable states strongly violating an area law.}
In one-dimension, a cluster state vector is (with obvious adaptions at the boundaries) the unique eigenvector of a set of mutually commuting
{\it stabiliser operators}
\begin{equation}
	K^{(j)}= Z^{(j-1)}X^{(j)}Z^{(j+1)}
\end{equation}
for $j=2,\dots, n-1$. It is left as an interesting and not very difficult exercise to the reader to find out how this state vector can be represented 
as an MPS with local dimension $d=2$ and bond dimension
$D=2$.

In Ref.\ \cite{Models1} {\it new models for measurement-based computing} have been proposed, taking the idea seriously that 
the matrices used in the parametrization of an MPS can be directly understood as quantum gates on a logical space. Indeed, this mindset gives rise to a wealth of novel
models, an idea that has turned out to be fruitful since then. For example, resource states can be found exhibiting long-range correlations and variants of the ground state
of the AKLT model can be taken to be resource states \cite{Models2,AKLTModels}. In Ref.\ \cite{Wires}  
a complete classification of {\it qubit wires} (spin systems allowing for a transport of quantum information) is given in
an instance where a physically well-motivated class of universal resources can be fully understood, using ideas of classifications of quantum channels.

\subsubsection{Localizable entanglement}
The ideas of the previous subsection also relates to the concept of {\it localizable entanglement} \cite{Localisable}: This is the characteristic 
length scale with which two sites $A,B\in V$ can be entangled by making local projective measurements on all other sites $V\backslash\{A,B\}$.
This length scale can be much longer than the traditional correlation length as in Eq.\ (\ref{CorrelationLength}). In fact, there are
gapped quantum Hamiltonians the unique ground state of which exhibits an infinite localisable entanglement, but a finite correlation length.

\subsubsection{Matrix product states in quantum state tomography}

The same reason why MPS (and again more general tensor network states) are so powerful in numerical approaches to problems in condensed matter physics render them also
optimally suited for another purpose: For {\it quantum state (or process) 
tomography}. This is the important and natural task of estimating an unknown state (or process) from measurements. This 
is obviously one of the key tasks that experimentalists routinely face when performing experiments with precisely controlled quantum systems. It is beyond the 
scope of the present chapter to give a comprehensive overview over this important field of research. Still, from the above it should be clear where the insights 
developed here come in: In order to faithfully reconstruct an unknown pure generic many-body state of local dimension $d$ and $n$ sites from expectation values, one needs 
to know $O(d^{n})$ different expectation values. In order to reconstruct an unknown MPS, in sharp contrast, merely $O(n D^2 d)$ expectation values are needed,
an exponential improvement. What is more, one can obtain the relevant information from learning suitable {\it reduced density operators} alone \cite{EfficientTomo}.
Similar ideas can also applied to {\it quantum fields} and continuum systems, using the concept of continuous matrix product states that we will encounter later
\cite{Wick}. Without such tools, it seems that the certification of quantum experiments can soon no longer 
keep up with experimental progress with controlled quantum systems.
\index{Quantum state tomography}

\section{Higher-dimensional tensor network states}

The idea of a tensor network is by no means confined to one-dimensional quantum systems. In fact, one would usually rather refer to an actual `tensor network' if 
the topology is not that of a one-dimensional chain. We start with the higher-dimensional analogue of matrix product states and then turn to other approaches such
as multi-scale entanglement renormalisation.

\subsection{Higher-dimensional projected entangled pair states}

\subsubsection{Definition of projected entangled pair states}

A {\it projected entangled pair state} (PEPS) \cite{PEPSOld}, closely related to the older concept of a {\it tensor network state} \cite{Nishino,Sierra,Niggemann}, is 
the natural generalisation of an MPS to higher-dimensional systems. For a cubic lattice $V=L^{\cal D}$
for ${\cal D}=2$ and open boundary conditions, the tensor network can be graphically represented as
\begin{figure}[h!]
 \centering
\includegraphics[width=0.45\textwidth]{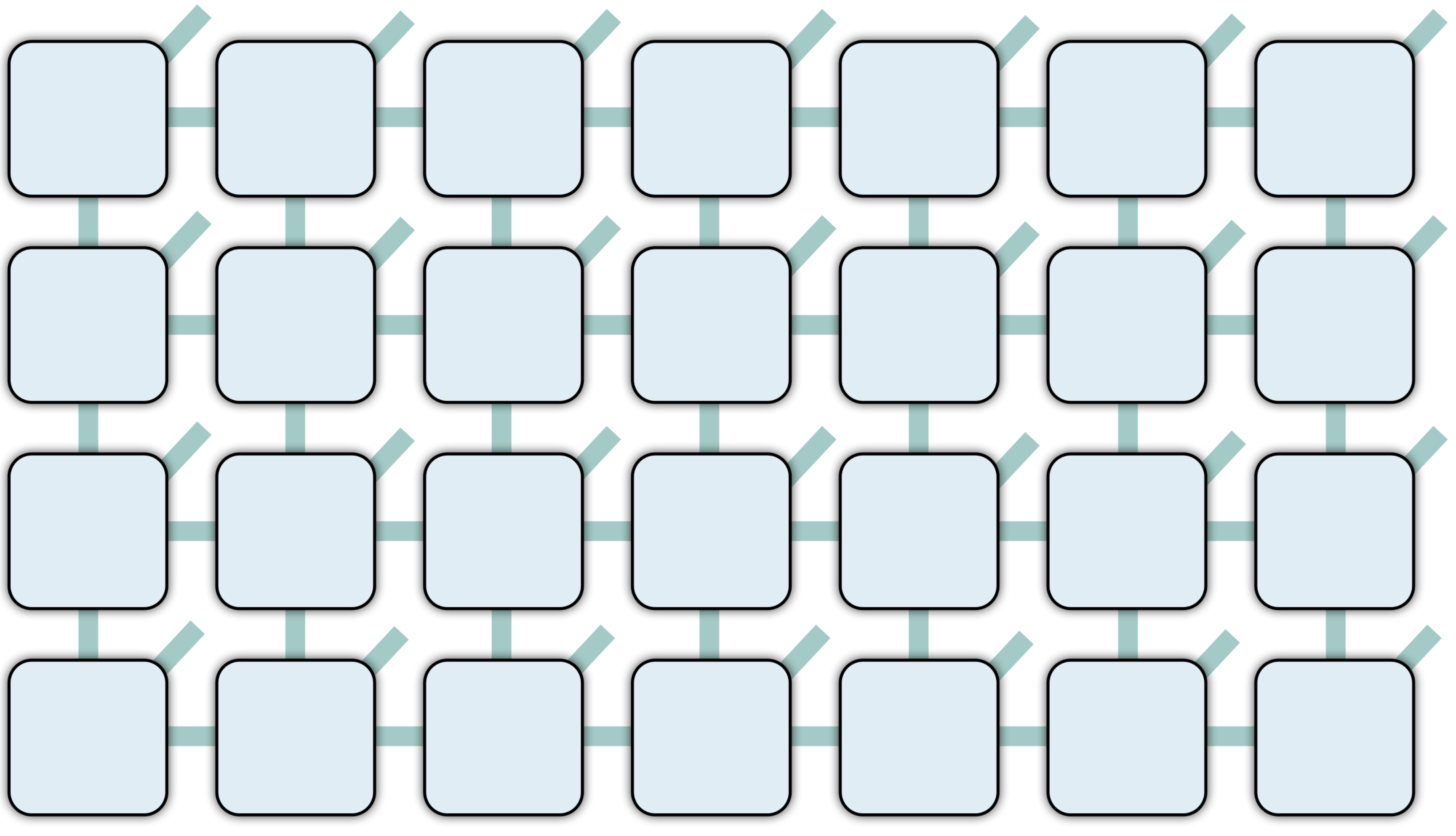},
\end{figure}
\index{Projected entangled pair state}

with the natural analogue for periodic boundary conditions on the torus.
Similarly as before, each of the tensors -- now five index tensors $A^{(k)}_{\alpha,\beta,\gamma,\delta;j}$ --
can be chosen all different for all sites $k\in V$, with $\alpha,\beta,\gamma,\delta=1,\dots, D$ and $j=1,\dots, d$.
But they can again also all the same (possibly with the exception of the tensors at the boundary) 
in a translationally invariant ansatz class. 

Again, one can formulate an actual projected entangled pair picture: Imagine again that each pair of physical sites in $V$ shares with each nearest neighbour 
a maximally entangled state as defined in Eq.\ (\ref{maxent}), in the virtual degrees of freedom. To this state again a linear map $P^{(j)}$ is applied for each site,
now (with variations at the boundary for open boundary conditions) a map $P^{(k)}:(\cc^D)^{\otimes 4}\rightarrow\cc^d$, defined as
\begin{equation}
	P^{(k)} = \sum_{\alpha,\beta,\gamma,\delta=1}^D \sum_{j=1}^d A^{(k)}_{\alpha,\beta,\gamma,\delta;j} |j\rangle\langle \alpha,\beta,\gamma,\delta|.
\end{equation}
Again, one `projects entangled pairs',
\begin{figure}[h!]
 \centering
\includegraphics[width=0.4\textwidth]{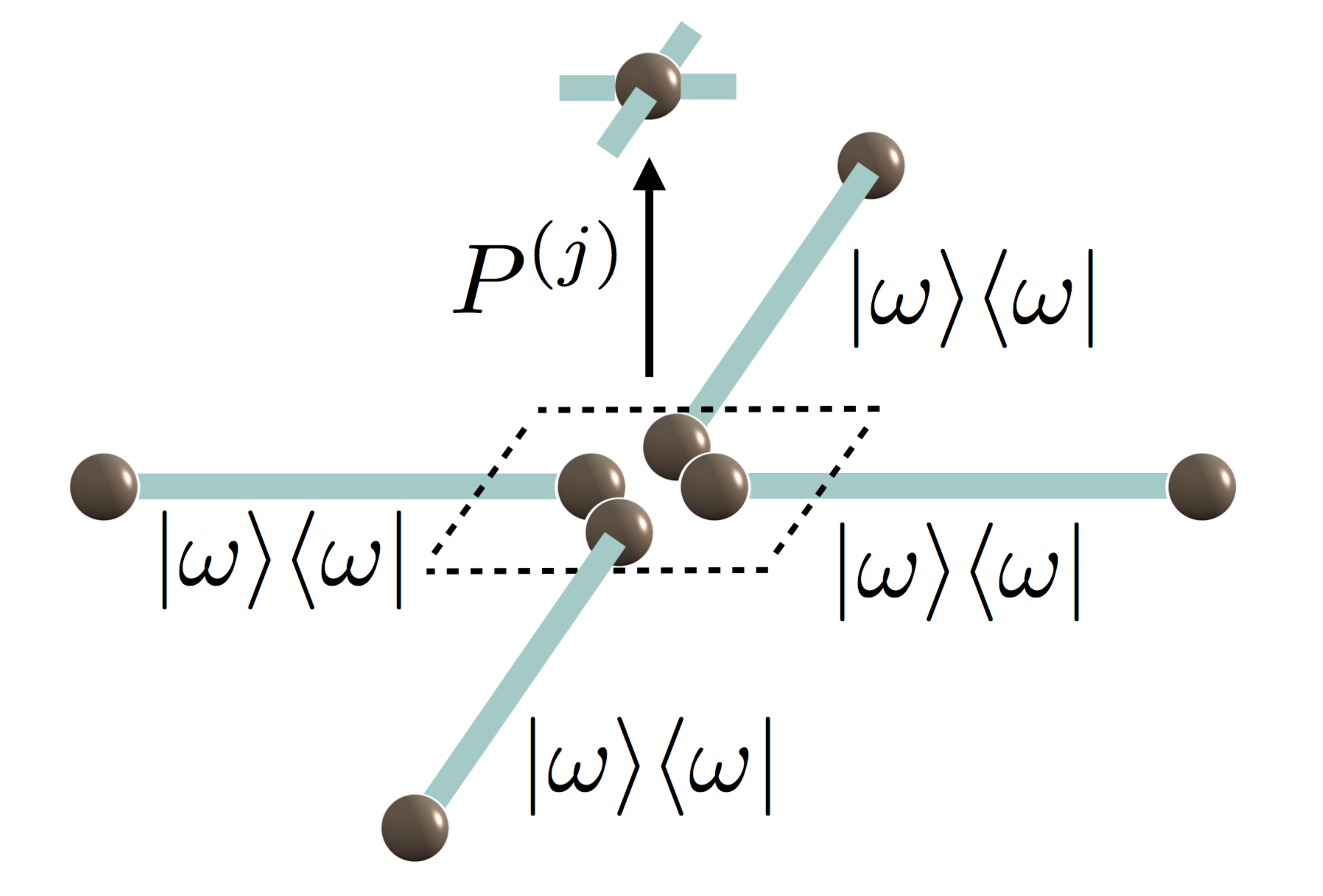}.
\end{figure}

\subsubsection{Properties of projected entangled pair states}
A number of natural properties emerge that render this class of variational states a natural one: 
\begin{itemize}
\item PEPS
satisfy area laws, and it takes a moment of thought to see that the entanglement entropy of a subset $A$ is bounded from above by $O(L \log D)$ for ${\cal D}=2$:
one simply needs to disentangle as many bonds as the boundary $\partial A$ of $A$ contains. 
\item Again, if the bond dimension $D$ is large enough, then one can approximate (or for finite systems explicitly write out) every state as a PEPS. 
\item One can also again have exponentially clustering of correlations. Interestingly, here already a
difference emerges to MPS in one dimension: One can construct PEPS that have algebraically decaying correlations with $\dist (A,B)$ between two sites or regions
$A,B\subset V$ \cite{PEPSArea}.
\end{itemize}
Such a strong statement on how well general states can be approximated with PEPS as it is available for MPS is lacking: One expects, however, that states satisfying area laws -- so 
presumably ground states of gapped models -- can be well approximated with PEPS with a small bond dimension $D$. Also, the body of numerical evidence available
shows that this class of states indeed meaningfully describes strongly correlated quantum many-body systems.

\subsubsection{Approximate contraction of projected entangled pair states}

Again similarly to the situation before, one can define {\it transfer operators} $E_\id$, graphically represented as
\begin{figure}[h!]
 \centering
\includegraphics[width=0.28\textwidth]{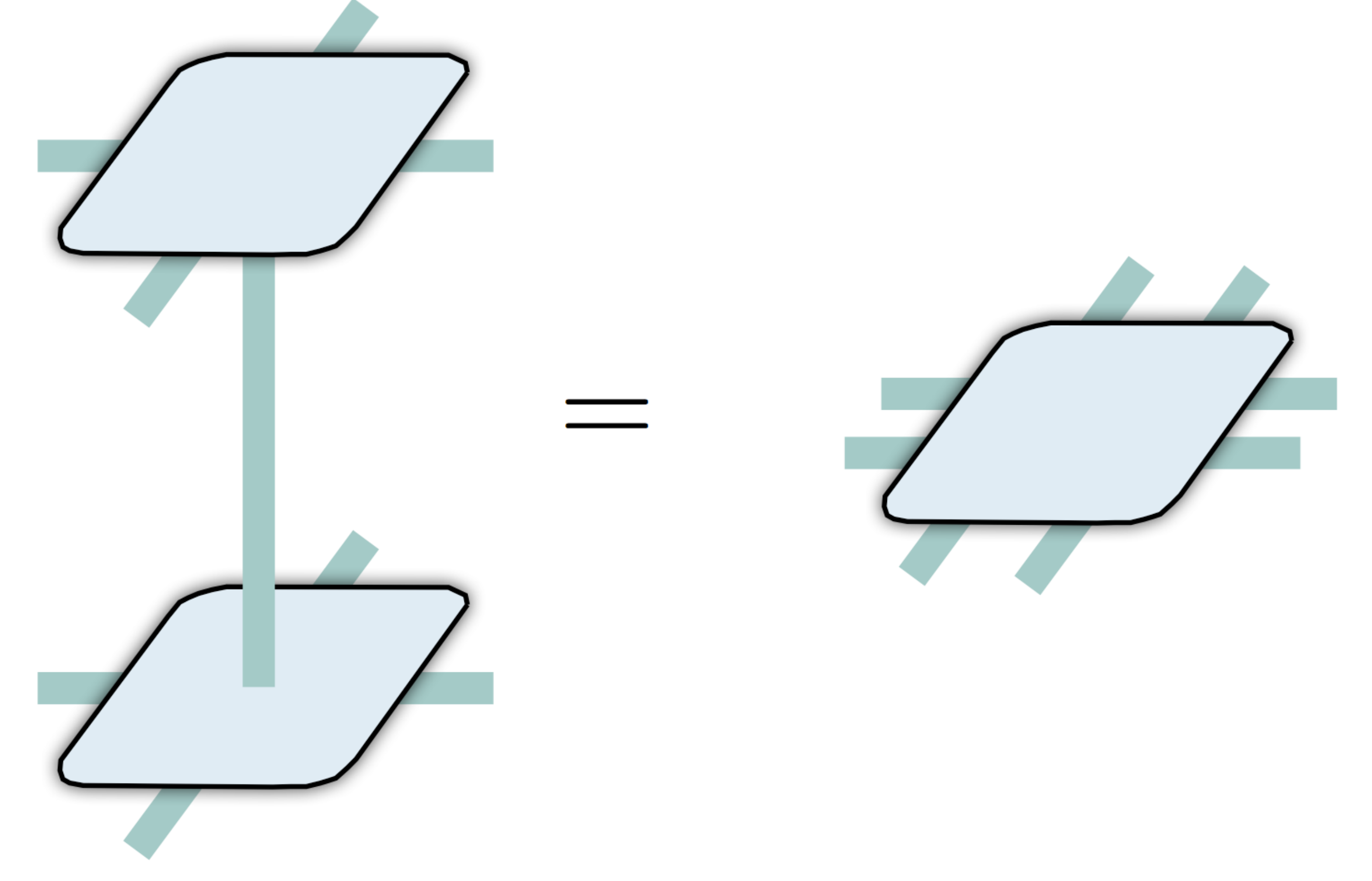}
\end{figure}

and similarly $E_{O_A}$ in case of the presence of a local observable.
In contrast to MPS, PEPS cannot be exactly efficiently contracted, however.  Strictly speaking, the contraction is contained in the complexity class $\# P$, again giving
rise to a computationally hard problem \cite{Contraction}. It is also intuitive why naive exact contractions cannot work: No matter what contraction order one picks, one will end 
up with an object having a number of edges that is linear in $L$.
Therefore, computations such as the determination of expectation values of observables $O_A$ are, different from the
situation in MPS, strictly speaking inefficient. This is no reason for panic, however: There are promising results on {\it approximate contraction techniques} that allow
for a good practical description of gapped local two-dimensional Hamiltonians. For example, one can work oneself through the scheme and contract each row with the
subsequent one: Before progressing, however, one approximates the bond dimensions of the new tensors by those the tensors had previously \cite{PEPSOld}. 
This is no longer exact, but feasible and efficient (the best known such scheme displays an effort of $O(D^8)$ in the bond dimension). 
Another approximate way of contracting amounts to
renormalising the cubic lattice to a new cubic lattice of smaller system size, and again approximates the obtained tensors of higher bond dimension by the previous one
\cite{GuAdded,Jiang}. There has also been recent insight into why such approximation methods should be
expected to provide good service, using ideas of {\it local topological order} \cite{Robustness}.

%
%

\subsubsection{Methods for infinite lattices}

Again, the basic ideas have been generalised to the situation of having an infinite system to start with, to avoid the need for finite size scaling. 
A number of methods have been suggested, among them {\it boundary-MPS methods}
\cite{iPEPS}, {\it corner transfer matrix} methods  \cite{Nishino}, as well as again {\it tensor coarse-graining methods} \cite{GuAdded,Jiang}.
These methods provide competitive numerical simulations of two-dimensional lattice models, for a review, see Ref.\ \cite{Orus}.\footnote{To be fair, one should add that at present,
one-dimensional approaches based on matrix product states are still better developed than those based on 
higher-dimensional tensor networks. The `crime story'
of the precise nature of ground state of the spin-1/2 {\it Heisenberg anti-ferromagnet on the
Kagome lattice} with 
local Hamiltonian terms as in Eq.\ (\ref{Heisenberg}) -- it is a spin liquid ground state -- has finally
been resolved using DMRG and a
`snake-like' one-dimensional ordering of the tensors of the two-dimensional Kagome lattice, and not using
an algorithm using PEPS or multi-scale renormalisation \cite{Crime1,Crime2}.
}

\subsubsection{Exact models}

It goes without saying that again PEPS are not only a useful tool in numerical studies, but in analytical ones as well. {\it Cluster states} in two dimensions \cite{Oneway}
are instances of PEPS, and so are a number of other classes of states important in quantum information theory. The models of Refs.\ \cite{Models1,Models2} for {\it measurement-based
quantum computing} are also based on PEPS.
The possibly most important Hamiltonian with PEPS as ground states is the {\it toric code Hamiltonian}
\begin{equation}
	H = -J_a \sum_s A_s - J_b \sum_p B_p
\end{equation}
defined on the edges of a two-dimensional cubic lattice,
where $\{A_s\}$ and $\{B_p\}$ are {\it star} and {\it plaquette operators}, respectively, defined as
\begin{equation}
	A_s = \prod_{j\in s} X^{(j)},\,\,
	B_p = \prod_{j\in p}Z^{(j)},
\end{equation}
so by the product of Pauli operators around a star or around a plaquette. On the one hand, this model is interesting as it can be viewed as a lattice instance of a
$\zz_2$ {\it lattice gauge theory}. 
On the other hand, it is the most prototypical quantum lattice model exhibiting {\it topological order}. It may hence by no surprise that the
literature on this model is for good reasons very rich, to say the least.
\index{Toric code}

There are also models considered in the literature that can not be exactly solved by means of PEPS, but for which variational approaches of PEPS
with few variational parameters already give both good energies, but at the same time significant physical insight into the model at hand, so have a somewhat analytical
flavour. Such a mindset
has been followed, e.g., in the study of {\it resonating valence bond wave-functions} \cite{VBS}.

\subsection{Multi-scale entanglement renormalization}

So far, we have been discussing tensor networks that had the same topology as the underlying physical lattice. Needless to say, there are good reasons to 
choose other topologies as well. A guideline is served by the criteria that  
\begin{itemize}
\item[(i)] the tensor network should be described by polynomially many parameters, 
\item[(ii)]
it should be efficiently contractible, either exactly or approximately, and 
\item[(iii)] the corresponding class of quantum states 
should be able to grasp the natural entanglement or correlation structure.
\end{itemize}

\subsubsection{Tree tensor networks}

Specifically, for critical models, one would expect a scale invariant
ansatz class to be reasonable, one that reflects the scale invariant nature of the ground state. A first attempt in this direction is to think of 
tree tensor networks \cite{Tree1,Tree2}. For example, one can think of a {\it binary tree tensor network}: Here, one introduces a fictious time in a renormalisation scheme where
in each step, two sites are mapped into a single one by an {\it isometry}. At the top layer of this hierarchy, one prepares two systems in some initial pure state. This scheme
has several of the above 
advantages: (i) It is described by polynomially many parameters, (ii) one can efficiently contract the network, and (iii) the states generated inherit the scale invariance
of the ansatz class. There are also disadvantages: notably, there are sites in the physical lattice that are nearest neighbours; yet, in the tensor network they are only
connected via the top layer in the hierarchy.
Tree tensor networks with an entire symmetric subspace on the top layer and not only a single state naturally emerge as exact 
{\it ground states of frustration-free
spin Hamiltonians} \cite{FF}.

\subsubsection{Multi-scale entanglement renormalisation}

A generalisation has been suggested in Ref.\ \cite{MERA1}, referred to as {\it multi-scale entanglement renormalisation (MERA)}. The idea can be 
viewed as a refinement of the binary tree mentioned above, to what is called a binary MERA. For clarity of notation, we consider one-dimensional chains.
Again, one thinks of a tensor network having different {\it layers} or temporal steps. Let us imagine that $n=2^T$, meaning that we think of $T$ temporal layers,
labeled $t=1,\dots ,T$. $t=1$ corresponds to the physical layer, $t=T$ to the top layer.
However, a new ingredient is added. Each temporal step now consists of two elementary steps. One elementary step is again a
renormalisation step, invoked by an isometry $I:\cc^{d_j\otimes d_j}\rightarrow \cc^{d_{j+1}}$ 
satisfying $I^\dagger I=\id$ (with $d_1=d$) \index{Multi-scale entanglement renormalisation}
\begin{figure}[h!]
 \centering
\includegraphics[width=0.34\textwidth]{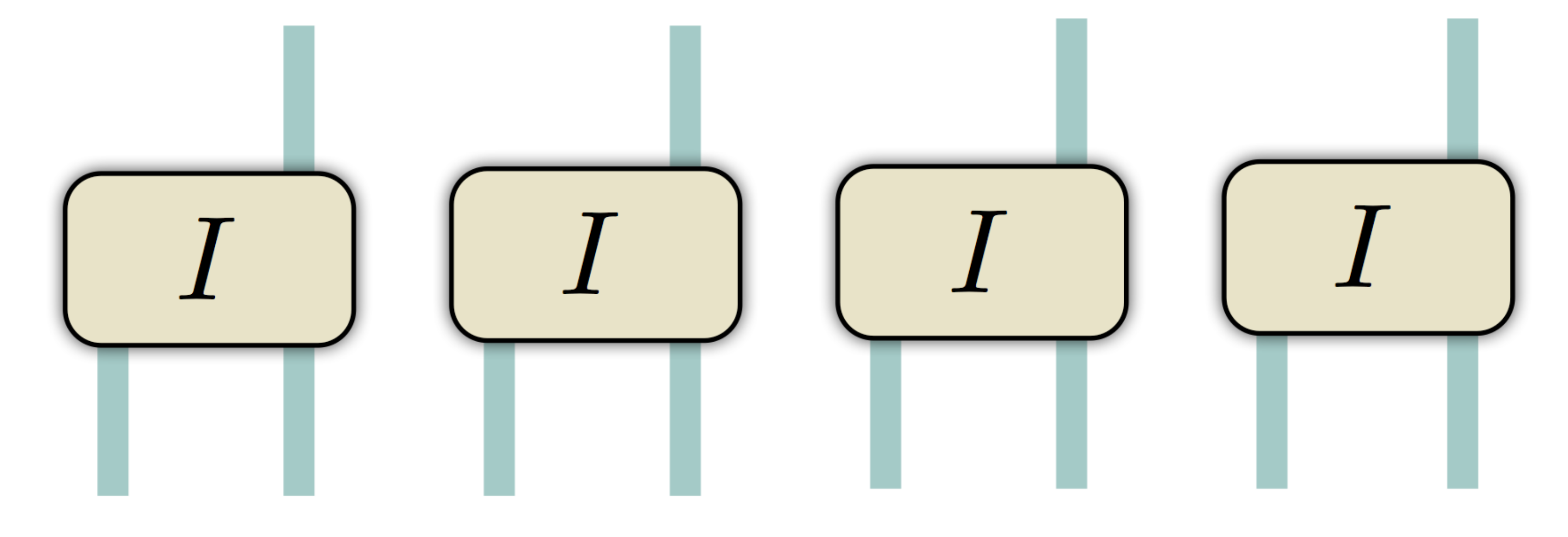}.
\end{figure}

In addition, an interlaced 
elementary step of a layer of {\it disentanglers} is introduced, so unitaries $U\in U(d_j^2)$ that let respective pairs of nearest neighbours 
interact,
again satisfying $U^\dagger U=\id$, 
\begin{figure}[h!]
 \centering
\includegraphics[width=0.34\textwidth]{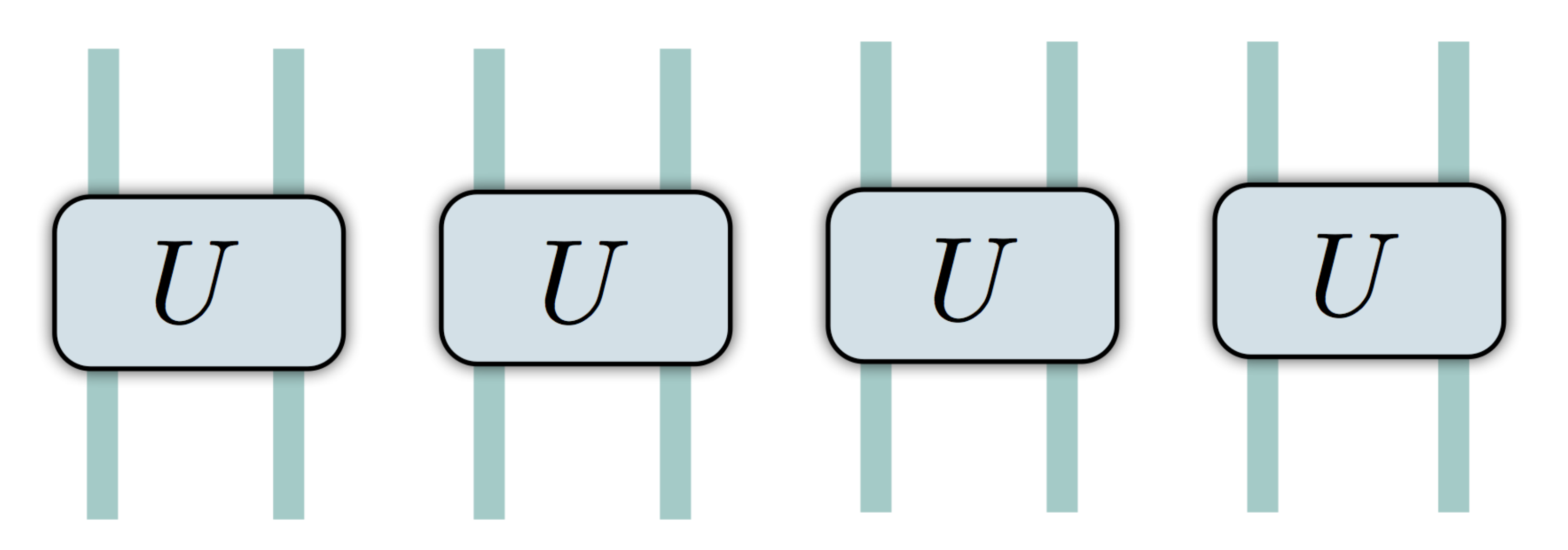}.
\end{figure}

The rationale behind this approach is that one first disentangled neighbours as much as possible 
before they are renormalised separately. (i) Again, one has polynomially many degrees of freedom, if $d_{\text{max}}=\max\{d_j\}$, actually 
$O(d_{\text{max}}^2 n \log(n))$ many real parameters. 
(ii) Then, contraction is still efficient. This might not be entirely obvious. The key insight is that when computing expectation values 
$\langle \psi|h_j |\psi\rangle$ for a Hamiltonian term $h_j$,
since all steps are either unitary or isometric,
one can remove all tensors outside the {\it causal cone} of the Hamiltonian term $h_j$, and the tensors within the causal cone can be sequentially
contracted following the temporal order: It is clear by construction that the causal cone will have a
finite width.
The subsequent picture depicts the tensor network of a MERA, with the causal cone of a Hamiltonian term being marked
with tensors having a slightly darker colour,
\begin{figure}[h!]
 \centering
\includegraphics[width=0.49\textwidth]{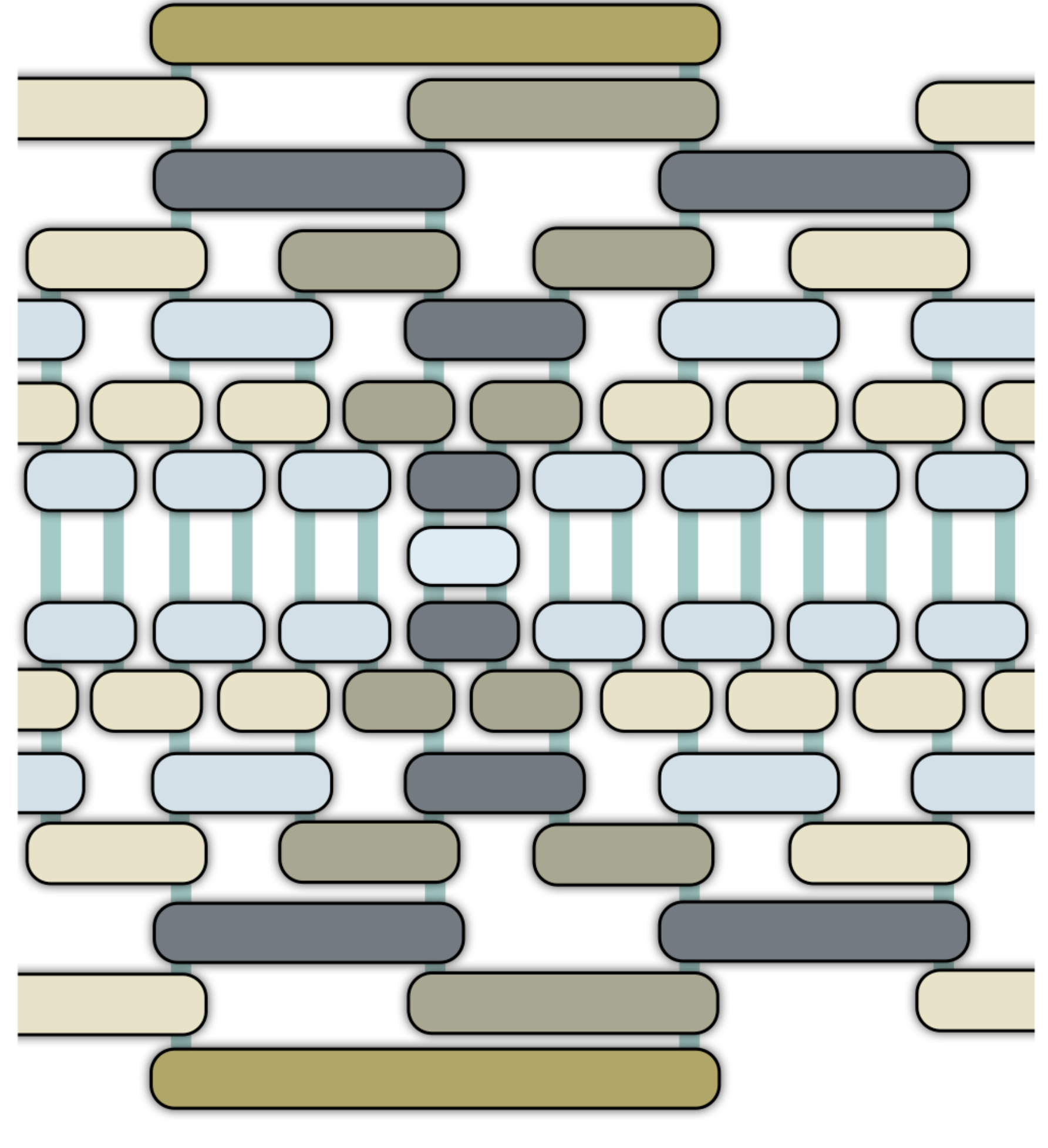}.
\end{figure}

Therefore, the contraction is possible with polynomial effort in $n$ and in $d_{\text{max}}$. This statement remains true for dimensions 
${\cal D}>1$: Hence, MERA constitute an efficiently contractible network in fact in any dimension. It turns out that for ${\cal D}>1$ and cubic lattices,
MERA can always be mapped onto a
PEPS of finite bond dimension, so the set of MERA states may be viewed as a subset of PEPS \cite{MeraIsPeps}. However, MERA can always be exactly efficiently 
contracted and they exhibit a very special and meaningful structure. (iii) Indeed, MERA can be put into contact with critical systems and {\it conformal field
theories}. In several ways, one can view a MERA as a lattice instance of a conformal field theory, an insight that has been fleshed out in quite some detail
\cite{MERAC2,MERAC1,Swingle}. First numerical 
implementations of this idea of a MERA were presented in Refs.\ \cite{Flow,Rizzi}; in 
the meantime, these approaches have also been generalised to higher-dimensional systems 
\cite{AlgorithmsMera,EvenblyNew,MERAF2,MERAF4}.

\section{Fermionic and continuum models}

\subsection{Fermionic models}

\subsubsection{Fermionic tensor networks}

Models that have {\it fermionic degrees of freedom} associated with the lattice sites can be mapped onto spin models with $\cc^2$ constituents 
by virtue of the {\it Jordan Wigner transformation}. However, only for one-dimensional models this approach is practical, as only then a local fermionic model is mapped 
again onto a
local spin model. For two-dimensional fermionic models on $L^{\cal D}$ with ${\cal D}=2$, say, in contrast,
one encounters highly non-local terms that have a locality region growing as $\Omega(L)$, no matter what specific ordering is chosen,
rendering the methods discussed here inappropriate (and the computation of expectation values inefficient). This seems unfortunate, in particular since some of the
challenging and at the same time most interesting models such as the two-dimensional fermionic {\it Hubbard model} with Hamiltonian
\begin{equation}
  H=-t \sum_{\langle j,k\rangle,\sigma} (f_{j,\sigma}^\dagger f_{j,\sigma}+ \text{h.c.})
  + U \sum_j f^\dagger_{j,\uparrow}  f_{j,\uparrow} f^\dagger_{j,\downarrow}  f_{j,\downarrow} 
\end{equation}
for $t,U\in \rr$ are exactly of this kind.
\index{Fermionic models}
\index{Hubbard model}

There is a way to overcome this obstacle, however. The prejudice to be overcome is that one should not fix a global order beforehand, but rather update the local
order `on the fly', in a way such that all expectation values $\langle O_A\rangle$ are correctly reproduced and give the same values as if one had (inefficiently)
mapped the system onto a non-local spin model. 
This idea has been introduced in Refs.\ \cite{MERAF1,MERAF2,MERAF3,MERAF4} and further applied in Refs.\
\cite{MERAF5,MERAF6}.
One way of formulating the problem is in terms of {\it fermionic operator circuits}, 
more general objects than
standard tensor networks that also specify information about the local order of edges, correctly accounting for the signs encountered when {\it swapping fermionic modes}.

\subsubsection{Tensor networks in quantum chemistry}

An emergent field of research is the study of systems of {\it quantum chemistry} with tensor network methods.
These are again interacting fermionic models, but this time with a Hamiltonian that is lacking an obvious structure of locality.
In second quantisation, Hamiltonians in quantum chemistry can be written as
\begin{equation}
	H = \sum\limits_{i,j\in V}T_{i,j}f_i^\dagger f_j + \sum_{i,j,k,l\in V}
	V_{i,j,k,l}f_i^\dag f_j^\dag f_k f_l
\end{equation}
where $V$ and $T$ are some tensors that do not necessarily reflect geometrically local interactions of spinless fermionic modes. 
Yet, one can still order the orbitals suitably and consider this now as a 
 one-dimensional quantum system, albeit one with intricate interactions, and run a DMRG algorithm  \cite{Oers0,Oers}. 
 Also, one can employ tree-tensor network and complete graph approaches \cite{Troyer}. 
 It is an interesting emerging field to make use of tensor network ideas to capture such 
 models of quantum chemistry.

\subsection{Continuum models}

\subsubsection{Quantum fields}
We finally very briefly sketch how tensor network methods can be applied to capture continuous systems, as they arise in 
quantum field theory; obviously, we will not be able
to be fair to this subject, but rather provide a glimpse into it. The type of systems that can most naturally be captured in the way described below 
is constituted by
one-dimensional systems of bosons or fermions on line segments of length $L$, associated with {\it field operators}
$\Psi(x)$ and $\Psi^\dagger (x)$, where
\begin{equation}
	[\Psi(x) , \Psi^\dagger (y)]= \delta(x-y),\,\,
	\{ \Psi(x) ,\Psi^\dagger (y) \}= \delta(x-y)
\end{equation}
for bosons and fermions, respectively, for $x,y\in [0,L]$. A common Hamiltonian in this context is the {\it Lieb-Liniger model},
a non-relativistic model of a continuous bosonic system with a contact interaction, with Hamiltonian
\begin{equation}
	H = \int_0^L dx \left(
	\frac{d\Psi^\dagger(x)}{dx}
	\frac{d\Psi(x)}{dx}
	 + c 
	\Psi^\dagger (x) 
	\Psi^\dagger (x) 
	\Psi  (x)
	\Psi (x) 
	\right),
\end{equation}
for some $c>0$. This model is Bethe-integrable and physically important -- it models, for example, the situation of interacting cold bosonic atoms 
on top of atom chips -- so serves as a meaningful benchmark.
\index{Quantum fields}

\subsubsection{Continuous matrix product states}

{\it Continuous matrix product states} are variational states meaningfully describing such systems. Their state vectors are defined as
\begin{equation}\label{cMPS}
	|\psi\rangle = \tr_2\left(
	{\cal P}
	\exp\left(
	\int_0^L dx (Q(x) \otimes \id + R(x)\otimes \Psi^\dagger (x))
	\right)
	\right)|\o\rangle.
\end{equation}
where $|\o\rangle $ denotes the vacuum, and $\{R(x)\in \cc^{D\times D}: x\in [0,L]\}$ and 
$\{Q(x)\in \cc^{D\times D}: x\in [0,L]\}$ are position-dependent matrices reflecting a bond dimension $D$ \cite{cMPS1,cMPS2,cMPS3}, and ${\cal P}$
denotes path-ordering.
How can they seen to reflect  
the meaningful continuum limit of MPS?

 In the translationally invariant setting -- giving for simplicity of notation
rise to the situation that $R$ and $Q$ are not position dependent -- 
one can think for a given $L>0$ of $n=L/\varepsilon$ lattice sites
with infinite-dimensional
local Hilbert spaces of bosonic modes, and consider the limit $n\rightarrow  \infty$. For a given bond dimension $D$ and for $R,Q\in \cc^{D\times D}$ 
one can take as the matrices of the family of translationally invariant MPS
\begin{equation}
	A_1= \id + \varepsilon Q,\,\,
	A_2 = \sqrt{\varepsilon} R,\,\,
	A_k = \sqrt{\varepsilon}^k R^k/k!,
\end{equation}
for $k\geq 1$,
and identify $\Psi_j = a_j/\sqrt{\varepsilon}$ for $j=1,\dots, n$, which indeed yields Eq.\ (\ref{cMPS}) in the limit $n\rightarrow\infty$. 
Again, in order to compute expectation values of polynomials of
field operators, one does not have to perform computations in the physical Hilbert space, but merely in the correlation space of the (continuous) 
MPS. Equilibrium properties of a one-dimensional quantum field are found to relate to non-equilibrium properties of a zero-dimensional
quantum field \cite{cMPS1,cMPS2,cMPS3}, in fact by integrating a Markovian master equation that has $R$ as its Lindblad
operator.

For example, density-density correlation functions of second order can be computed as
\begin{equation}
	\langle \Psi^\dagger(x) \Psi^\dagger(0) \Psi(0)\Psi(x)\rangle = \tr (e^{T(L-x)}(R\otimes \bar R) e^{Tx} (R\otimes \bar R), 
\end{equation}
where $T:= Q\otimes \id + \id\otimes \bar Q + R\otimes \bar R$ (which can be viewed as the Liouvillian governing the Markovian
master equation),  \begin{figure}[h!]
 \centering
\includegraphics[width=0.5\textwidth]{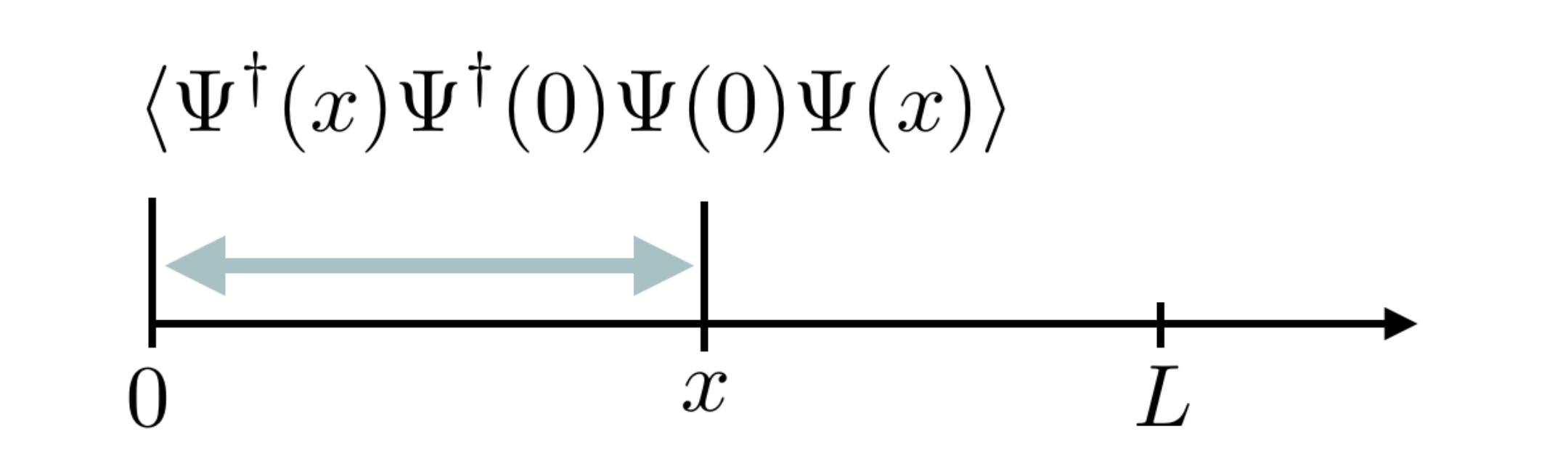}.
\end{figure}

Similarly, other correlation functions can be calculated from $R$ and $Q$ alone. This gives rise to a versatile
variational framework to capture continuous systems in a similar fashion as lattice systems are being grasped by matrix-product states.

\index{Continuous matrix product states}

%


%

\bigskip\bigskip

{\it Acknowledgements:} These lecture notes have profited from numerous fruitful discussions
with many people working on 
tensor network states in the past, most of whom 
appear in the list of references. I would specifically like to thank C.\ Krumnow, 
M.\ von Hase,
and E. Bergholtz for detailed feedback on the
manuscript. This work has been supported by the EU (Q-Essence, SIQS, RAQUEL), and the ERC (TAQ).

\bigskip
{\small
\providecommand{\bysame}{\leavevmode\hbox to3em{\hrulefill}\thinspace}
\providecommand{\MR}{\relax\ifhmode\unskip\space\fi MR }
\providecommand{\MRhref}[2]{%
  \href{http://www.ams.org/mathscinet-getitem?mr=#1}{#2}
}
\providecommand{\href}[2]{#2}

}

\end{document}